\documentclass[]{spie}  %>>> use for US letter paper
\usepackage{amsmath,amsfonts,amssymb}
\usepackage{graphicx}
\usepackage[colorlinks=true, allcolors=blue]{hyperref}
\usepackage{booktabs}
\usepackage{multirow}
\usepackage{color, colortbl}
\usepackage{xcolor}
\definecolor{Gray}{gray}{0.92}

\title{Per-clip adaptive Lagrangian multiplier optimisation \\ with low resolution proxies}

\author{Daniel J. Ringis}
\author{Fran\c{c}ois Piti\'e}
\author{Anil Kokaram}
\affil{Sigmedia Group, Electronic and Electrical Engineering Dept., Trinity College Dublin, Ireland}

\authorinfo{For further information please contact D.J. Ringis :  ringisd@tcd.ie}

\begin{document} 
\maketitle

\begin{abstract}
Optimising  the parameters of a video codec for a specific video clip has been shown to yield significant bitrate savings. In particular, per-clip optimisation of the Lagrangian multiplier in Rate controlled compression, has led to BD-Rate improvements of up to 20\% using HEVC. Unfortunately, this was computationally expensive as it required multiple measurement of rate distortion curves which meant in excess of fifty video encodes were used to generate that level of savings. 
This work focuses on reducing the computational cost of repeated video encodes by using a lower resolution clip as a proxy. Features extracted from the low resolution clip are then used to learn the mapping to an optimal Lagrange Multiplier for the original resolution clip.  In addition to reducing the computational cost and encode time by using lower resolution clips, we also investigate the use of older, but faster codecs such as H.264 to create proxies. This work shows the computational load is reduced by up to 22 times using 144p proxies, and more than 60\% of the possible gain at the original resolution is achieved. Our tests are based on the YouTube UGC dataset, using the same computational platform; hence our results are based on a practical instance of the adaptive bitrate encoding problem. Further improvements are possible, by optimising the placement and sparsity of operating points required for the rate distortion curves. Our contribution is to improve the computational cost of per clip optimisation with the Lagrangian multiplier, while maintaining BD-Rate improvement.

\end{abstract}

% Include a list of keywords after the abstract 

\section{INTRODUCTION}
\label{sec:intro}  % \label{} allows reference to this section

The amount of user generated video content has increased significantly \cite{wang2019youtube} to the point that it represents almost 80\% of all internet traffic \cite{cisco}. Hence video compression for video streaming applications has become even more important in the last decade since the development of H.264. Recent efforts in the community have led to the rapid deployment of next generation codecs e.g. H.265 (HEVC) \cite{sullivan2012overview} and VP9 \cite{mukherjee2013latest} and these are projected to displace H.264 in devices and browsers in the near future \cite{zhang2019overview}. Further iterations of these codecs already exist (H.266/VVC and AV1), but they are not yet in widespread use. This paper concentrates on performance of H.265 and VP9.

The key challenge facing any codec is the trade-off between rate and distortion. The accepted approach to controlling this trade-off is to express both terms in a cost \(J\) as follows. 
\begin{equation}
    J = D + \lambda R \label{rd}
\end{equation}
\(J\) combines both a distortion \(D\) (for a frame or macroblock) and a rate \(R\) (the number of coded bits for that unit) through the action of the Lagrangian multiplier \(\lambda\). The Lagrange multiplier approach was advocated by Sullivan et al \cite{sullivan1998rate} and that has been the adopted view in codec implementations since 1998. Any adjustment to the Lagrangian multiplier would lead to a change in the Rate-Distortion trade-off. This has an impact throughout the codec, as the idea is applied to many internal operations  e.g.  motion vectors, block type (Skipped/Intra/Inter), and bit allocation at the frame and clip levels \cite{wiegand2001lagrange}. The choice of \(\lambda\) was determined experimentally in the early days of codec research \cite{ortega1998rate}. A small dataset (less than 5 videos) was used to determine this potentially optimal choice of \(\lambda\). This small dataset may have been an adequate representation of videos at that time (1990s) but may not be ideal with the large range of user generated video content today.

\subsection{Contributions in this paper}

This paper explores the idea that a better $\lambda$ exists for an individual video clip which improves the BD-Rate compared to the current defaults used in existing implementations of the codecs considered. 

To generate results which are meaningful in modern day content at high resolutions, our first contribution is establish a reference of possible improvements for both VP9 and x265 on a large dataset of approximately 10,000 UGC video clips, by direct optimisation of $\lambda$ per clip. 

Second, as computational complexity is of key importance given the amount of video data being processed, we explore the use of proxy videos (different codec, different resolution, different codec settings) for estimating the optimised $\lambda$ at higher resolutions.

In addition to these direct optimisation schemes, we investigate the use of Machine Learning methods to directly predict the optimal $\lambda$ from the bitstream features.  

Results show up to 25\% gain in BD-Rate using direct optimisation, and up to 13\%  with Machine Learning methods specifically Random Forests. The next section presents a more detailed background and a review of recent work.

\section{Background}

The rate distortion algorithm \cite{wiegand1996rate} establishes a balance between the quality of the media and the transmission or storage capacities of the medium. As mentioned above, seminal work of Sullivan and Wiegand \cite{sullivan1998rate} laid the foundation for an empirical approach to choosing an appropriate \(\lambda\). This work established a relationship between the quantisation step size \(Q\) and the distortion \(D\) in a frame. Through minimising $J$, this leads to  a relationship between \(\lambda\) and \(Q\) expressed as   $ \lambda = 0.85 \times Q^2 $. A similar approach is deployed in the rate distortion algorithm for VP9. Which has a RD Cost multiplier directly proportional to $Q^2$ depending on the frame type \cite{vp9webm}. 
Updates to those experiments then yielded similar relationships for H.264 and H.265 (HEVC).  Because of the introduction of bi-directional (B) frames the constants are all different and three different relationships were established for each of the Intra (I), Predicted (P) and B frames as follows.
\begin{align}
   \lambda_{I} & =( 0.57 )2^{(Q-12)/3} \nonumber \\
\lambda_{P} & = (0.85) 2^{(Q-12)/3} \nonumber \\
\lambda_{B} & = ( 0.68 ) \max(2, \min(4, (Q - 12)/6))  2^{(Q-12)/3} \nonumber
\end{align}

\subsection{Adaptive Rate Control}

There exists a limited amount of work on adaptation of $\lambda$ in the rate distortion equation. In all cases, results support the idea that adjusting  \(\lambda\) leads to improvement in codec performance {\em per clip}. The idea is typically to adjust $\lambda$ away from the codec default by using a constant $k$ as follows.
\begin{equation}
   \lambda_{\textrm{new}} = k \times \lambda_{\textrm{orig}}\label{kfactor}
\end{equation}
where $\lambda_{\textrm{orig}}$ is the default Lagrangian multiplier estimated in the video codec, and  $\lambda_{\textrm{new}}$ is the updated Lagrangian.
The algorithms for estimating $k$ can be classified as Online and Offline.  In Online processing, $k$ is updated continuously during encoding, usually per GOP. In contrast, Offline processing is effectively a two pass process requiring an initial analysis stage followed by the use of a modified encode. 

\vspace{0.5em}\noindent {\bf Offline: } Ma et al \cite{ma2016adaptive} used a Support Vector Machine to determine \(k\). The {\em perceptual} feature set included scalars representing Spatial Information and Temporal Information, and a texture feature exploiting a Gray Level Concurrence Matrix. Details can be found in \cite{ma2016adaptive}. Their focus was on {\em Dynamic textures} and they used the DynTex dataset\cite{dyntex} of 37 sequences. They reported up to 2dB improvement in PSNR and 0.05 improvement in SSIM at equal bitrates. Hamza et al \cite{hamza2019parameter} also take a classification approach but using gross scene classification into indoor/outdoor/urban/non-urban classes. They then used the same $k$ for each class. 
Their work used the Derfs dataset\cite{derf} and reported up to 6\% BD-Rate improvement.

\vspace{0.5em}\noindent {\bf Online: } Zhang and Bull  \cite{zhang_bull} used a single feature ${D}_{P}/{D}_{B}$, the ratio between the MSE of P and B frames. This feature gives some idea of temporal complexity. Experiments based on the DynTex database yielded a choice for $k$ as follows.
\begin{equation}
    k = a({D}_{P}/{D}_{B}+d)^b +c\label{zhangbull}
\end{equation}
where \(a, b, c \textrm{ and } d \) were experimentally calculated and different for each codec tested (H.264 and H.265). In their work, $k$ was updated every GOP and they report up to 11\% improvement in BD-Rate. They modified $\lambda$ implicitly by adjusting the quantiser parameter $Q$. Papadopoulos et al\cite{Papadopoulos} exploited this and applied an offset to Q, in HEVC, based on the ratio of the distortion in the P and B frames. Each QP was updated from the previous Group of Pictures (GOP) using $ \textrm{QP} = a \times ({D}_{P}/{D}_{B}) - b $
where $a,b$ are constants determined experimentally.  This lead to an average BD-Rate improvement of 1.07\% on the DynTex dataset, with up to 3\% BD-Rate improvement achieved for a single sequence.

Yang et al \cite{yang2017perceptual} used a combination of features instead of just the MSE ratio above. In their work, they used a perceptual content measurement $S$ to model $k$ with a straight line fit $ k = aS - b$.
Here again $a,b$ were determined experimentally using a corpus of the Derfs dataset. The loss in complexity of the fit is compensated for by the increase in complexity of the feature. They report an average BD-Rate improvement of 6.2\%.

Recent work by Ling et al \cite{LingICME} indicated that the rate distortion performance of a video can be categorised. They have indicated that information about the video clips, in particular chrominance information, may lead to better RD-categorisation and subsequently improved RD-optimisation.

\subsection{Direct Optimisation of the Lagrangian Multiplier \label{dopt}}

As shown in equation \ref{kfactor} previous work attempts to adjust the rate control parameter $\lambda$ adaptively in some way. However there is no effort to establish a ground-truth {\em best possible} $k$. In our recently published work \cite{EIRingis} we introduce the use of direct optimisation to maximise BD-Rate improvement w.r.t $k$ for a given clip. This can be used as ground truth for developing a computationally more efficient approach with Machine Learning and/or proxies. %We repeat those experiments here, but for a much larger corpus and including VP9 as well. 
We present a summary of that previous work to set the context for our new contributions.

% In some way all these previous works have reported data showing $k$ versus $\textrm{BD-Rate}$, we can frame this such that $ \lambda_{\textrm{new}} = k \times \lambda_{\textrm{orig}}$. \textbf{eh??XXXACK}. Previous work uses various summary models to simplify this relationship across a corpus.

%There is no effort to establish a ground-truth {\em best possible} $k$. In this work we first use Direct Optimization methods to determine the upper bound for improvement of codec performance and hence yield a best possible $k$. Then this information is used to create a machine learning model to determine an appropriate value for \(k\) based on Offline video features. These ideas are presented next.

\begin{figure}
    \centering
    \includegraphics[width=0.4\columnwidth]{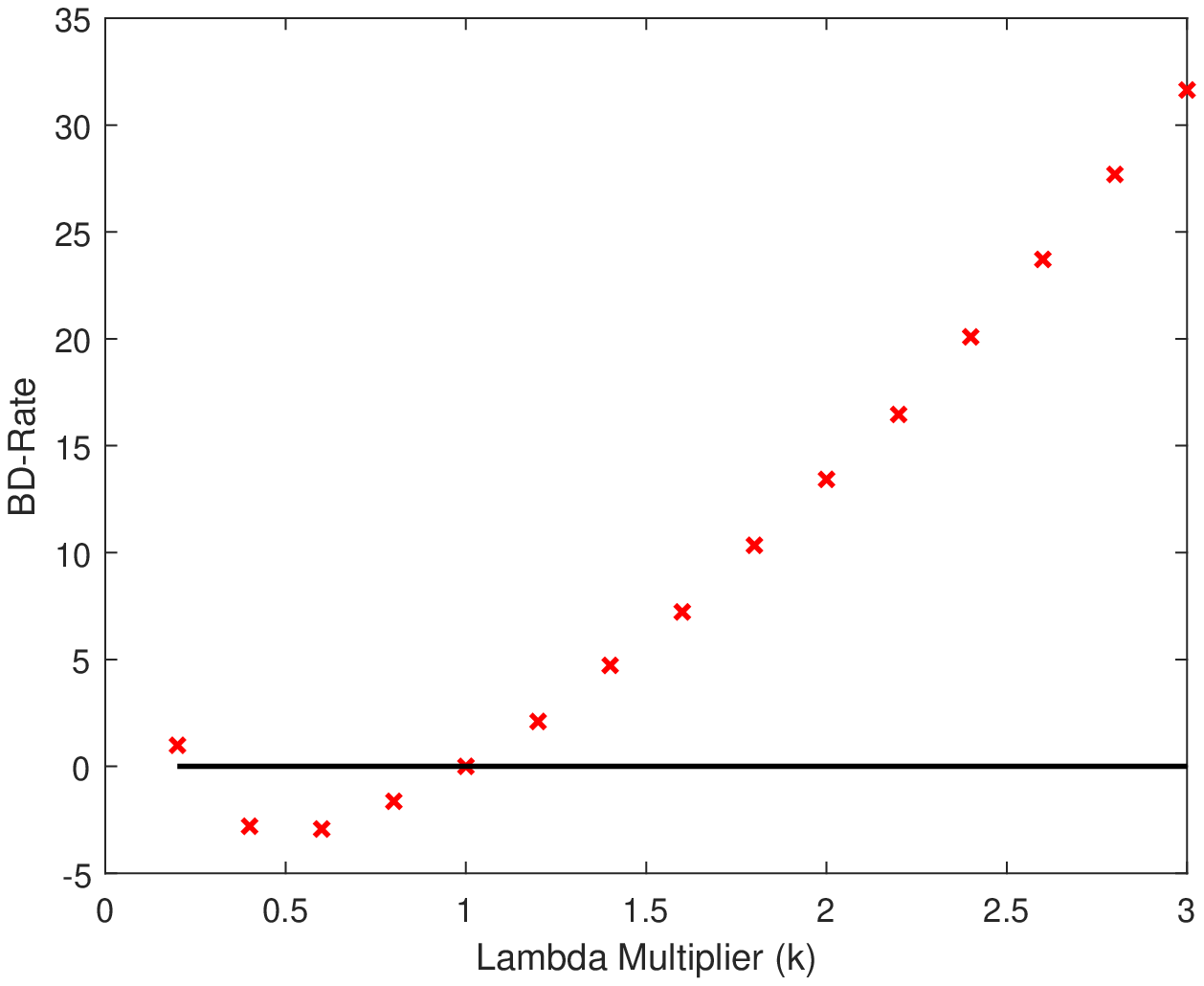}
    \includegraphics[width=0.4\columnwidth]{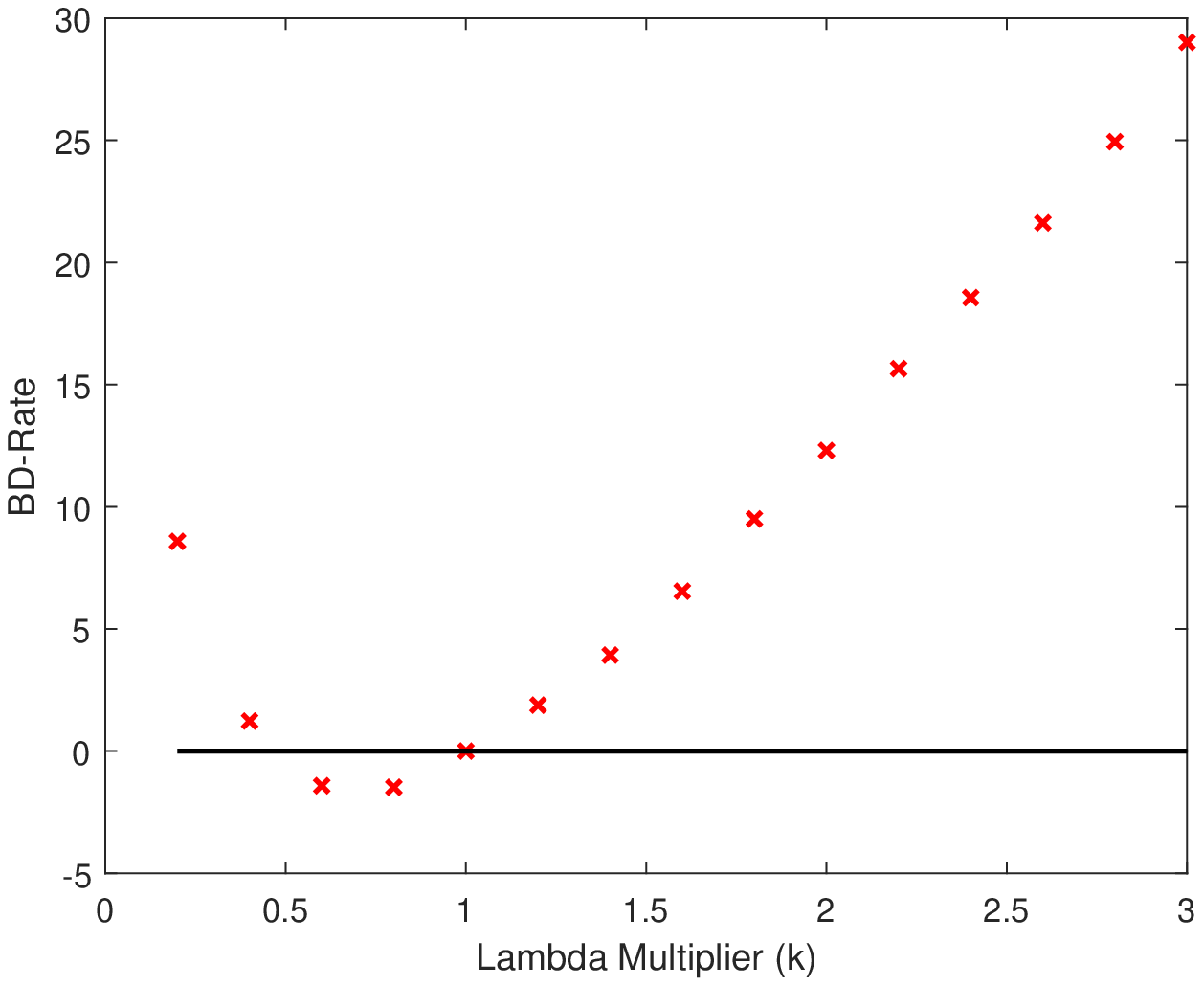}
    \\
    \includegraphics[width=0.4\columnwidth]{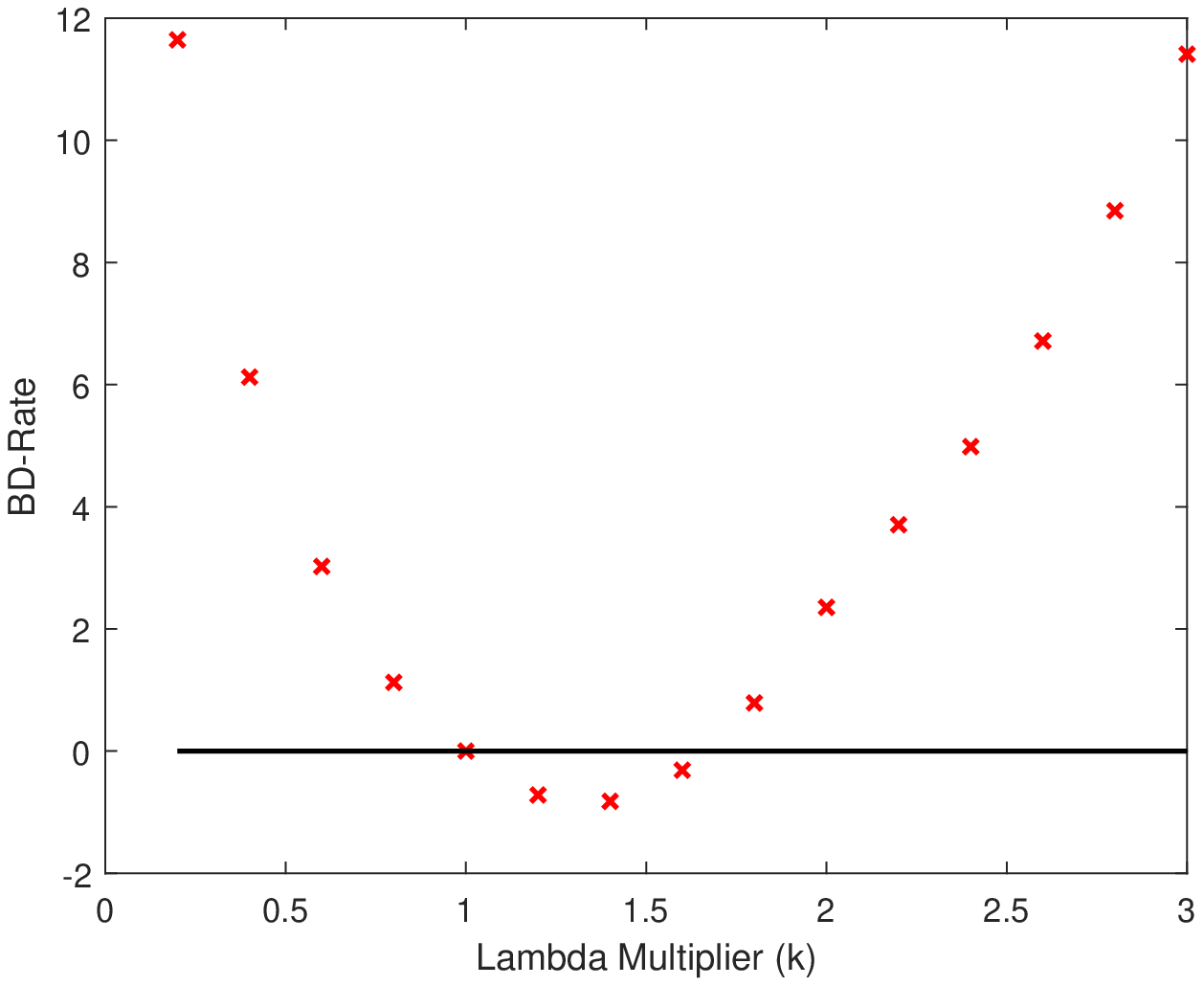}
    \includegraphics[width=0.4\columnwidth]{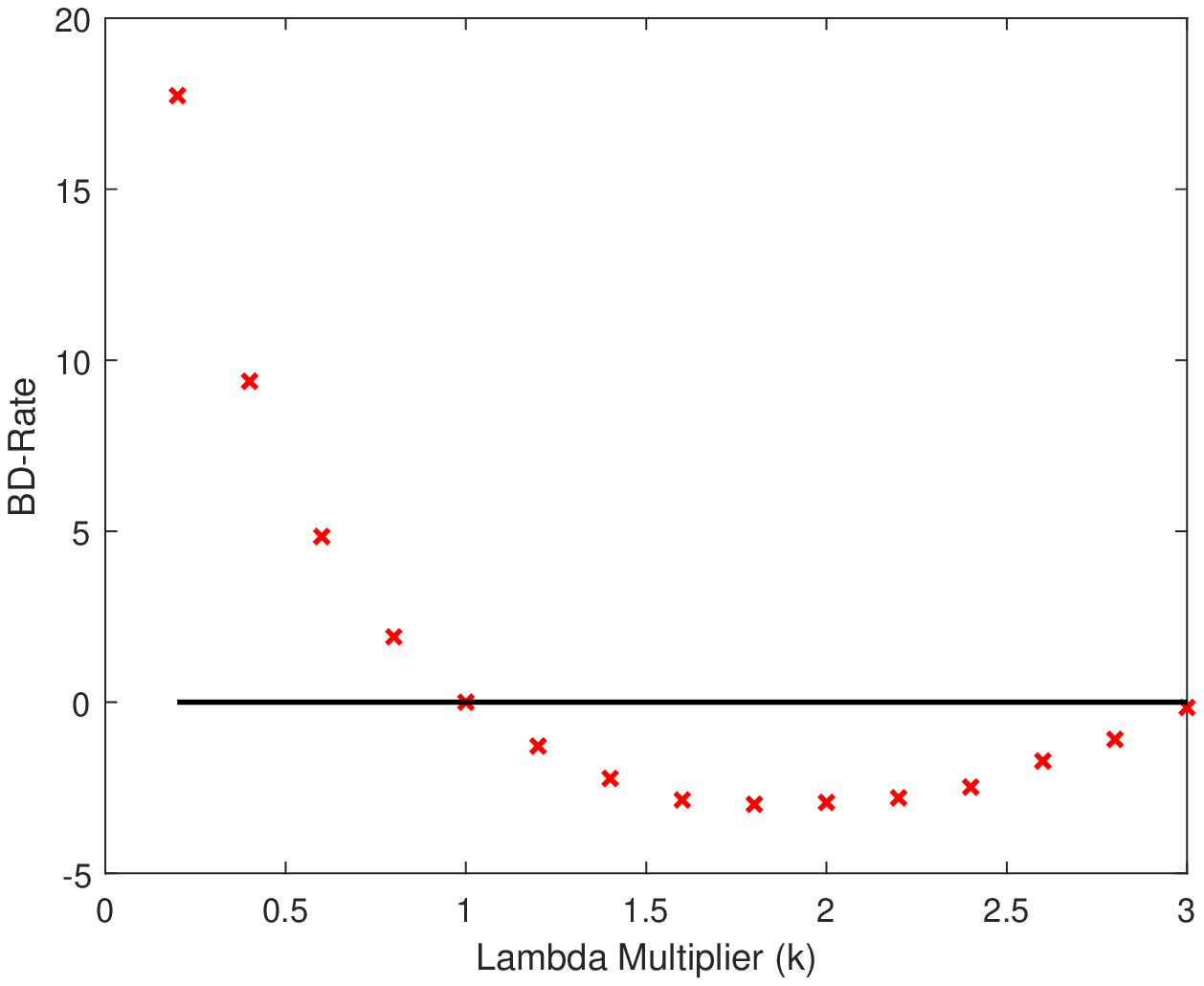}
    \caption{BD-Rate vs $k = [0.2:0.2:3]$ for four different clips (LiveMusic\_1080P-6d1a, MusicVideo\_480P-6fb6,  NewsClip\_720P-7e56 and Animation\_1080P-2fff). The best performing BD-Rate gain is achieved at a different value for $k$ in each clip. Hence per-clip optimisation is sensible. The curve shape shows a global minimum implying that classic optimisation strategies would be successful.}
    \label{exampleBDk}
\end{figure}

%For a given clip, we believe that there exists an ideal Lagrangian multiplier to provide the best possible bitrate while not affecting the quality of the video clip. 
In order to show that there is something to be gained in varying $k$ per clip, figure \ref{exampleBDk} shows BD-Rate vs $k$ for four different clips. There are two observations to be made. Firstly, the relationship between $k$ and BD-Rate varies substantially between clips. This reinforces the view that the traditional one size fits all approach is sub-optimal. Secondly, the general shape of the curves shows a global minimum. That minimum represents the maximum BD-Rate improvement available.

We directly minimise BD-Rate \cite{bdrate} with respect to $k$ ($B_r(k)$) using a scalar optimisation strategy, Brent's Method\footnote{Although any optimisation technique could be used.} \cite{numericalmethods}. We denote that estimate as $k_D$. The  objective function, $B_r(k)$, is therefore as follows.
\begin{equation}\label{bjontegaardEqn}
    B_r(k) = \int_{D_a}^{D_b} (R_1(D) - R_k(D)) dD
\end{equation}
Here the integral is evaluated over the quality range $D_a..D_b$.
$R_1(D), R_k(D)$ are the RD curves corresponding to the default ($k=1$) and the evaluated multiplier $k$ respectively. Each RD operating point is generated at a constant quality.  The overall optimisation process is then as follows.
\begin{enumerate}
    \item Generate an RD curve using $\lambda_{orig}$ and using 5 operating points (CRF 22,27,32,37,42). 
    \item Define our BD-Rate objective function as specified above in equation \ref{bjontegaardEqn}. We use the same polynomial-log fit for evaluating the integral as recommended by Bjontegaard\cite{bdrate}. PSNR is used for the Distortion criterion.
    \item Starting from $k=1.0$,  minimise  $B_r(k)$  wrt $k$ using Brent's method.\cite{numericalmethods}
\end{enumerate}
Note that for every evaluation of $B_r(k)$, five (5) encodes are required as well as the subsequent BD-Rate calculation itself. 
Optimisation was terminated when $B_r(k)$ reduced by less than 0.05\%.
 
\begin{figure}
    \centering

      \includegraphics[width=0.4\columnwidth]{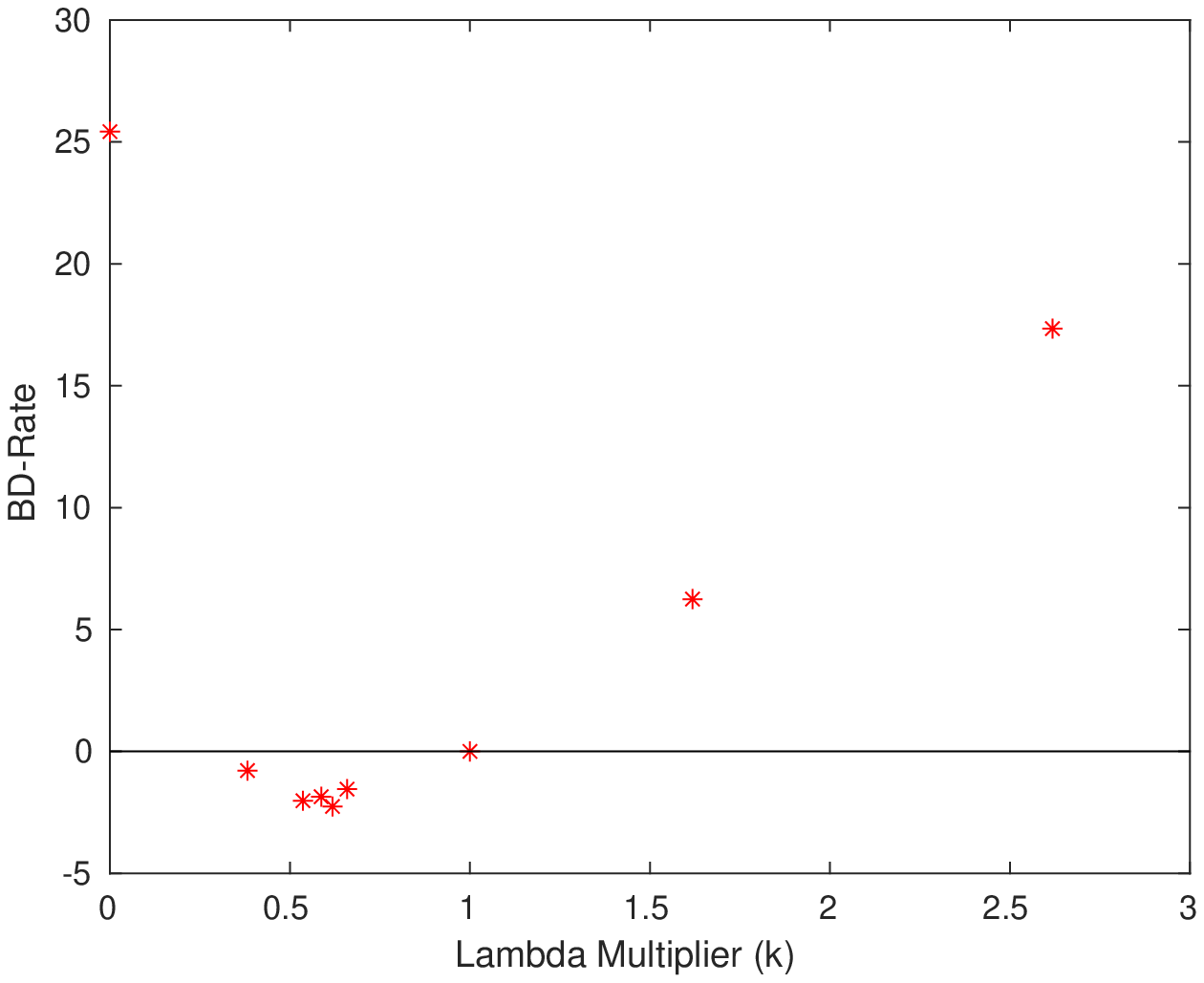} %\includegraphics[width=\columnwidth]{images/example11.eps}
    \includegraphics[width=0.4\columnwidth]{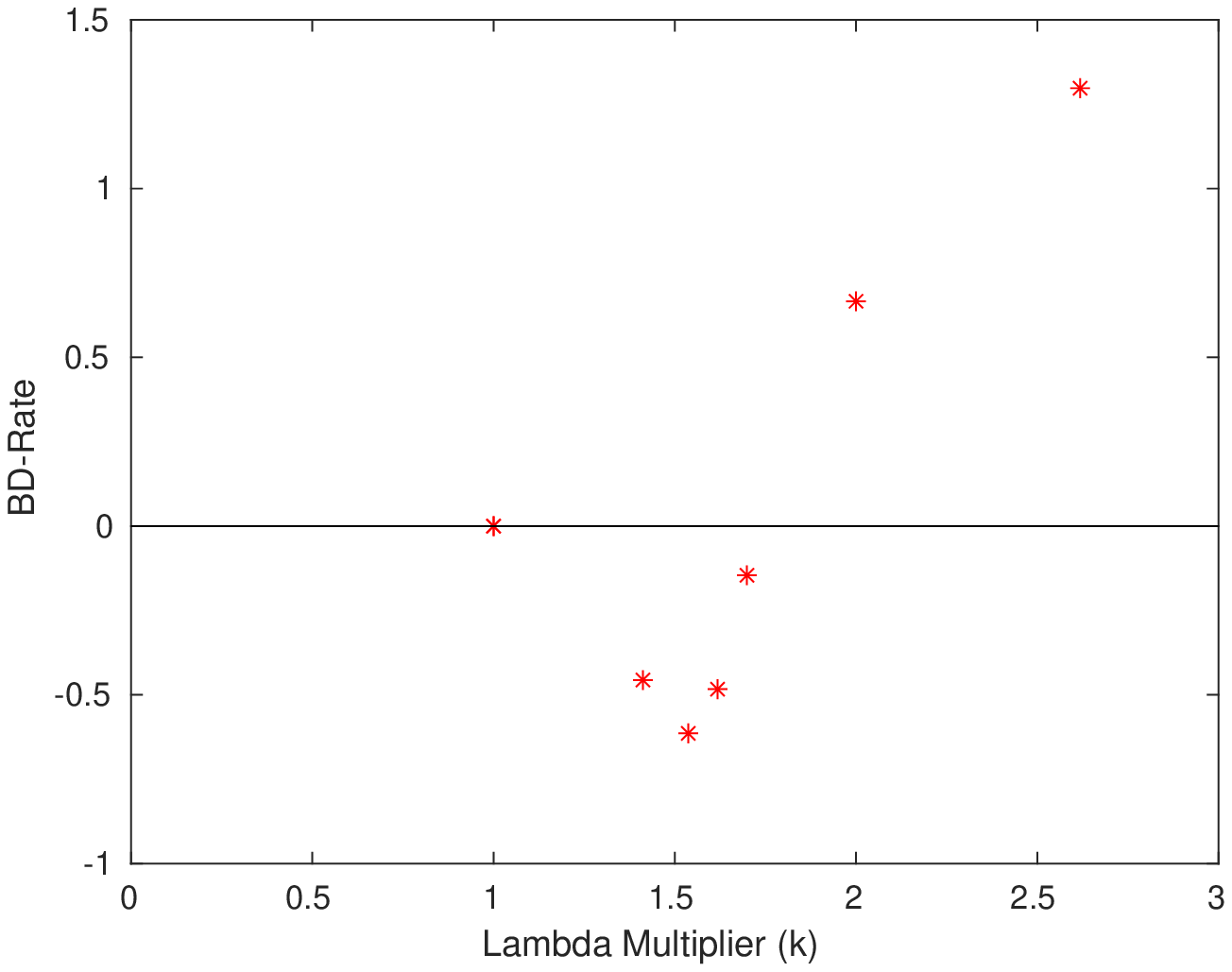}
    \caption{BD-Rate(\%) vs $k$ for two clips (Animation\_1080P-01b3 and  LiveMusic\_1080P-76fa). Only the points visited using the optmisation scheme (Brent's Method) are shown to give an idea that the process does converge. This leads to an improvement of 2.02\% and 0.61\% respectively.}
    \label{examplecurve}
\end{figure}

%  \begin{figure}
%      \centering
%      \includegraphics{BLOCK DIAGRAM EXPLAINING DIRECT OPTIMISATION ALGORITHM}
%      \caption{\textbf{BLOCK DIAGRAM EXPLAINING DIRECT OPTIMISATION ALGORITHM}}
%      \label{fig:my_label}
%  \end{figure}
%Brent's method \textbf{eh??XXXACK} was used as the  optimisation schemes\cite{numericalmethods} investigated. 
Figure \ref{examplecurve} shows the points visited in the optimisation search using Brent's method for two other clips (Animation\_1080P-01b3 and  LiveMusic\_1080P-76fa). This shows the optimisation being successful in each case, leading to a BD-Rate improvement of 2.02\% and 0.61\% respectively. In the next sections we use a much expanded corpus to generate  a statistically meaningful baseline of performance and then focus on computational complexity.

{}

\begin{figure}
    \centering
    \includegraphics[width=0.3 \columnwidth]{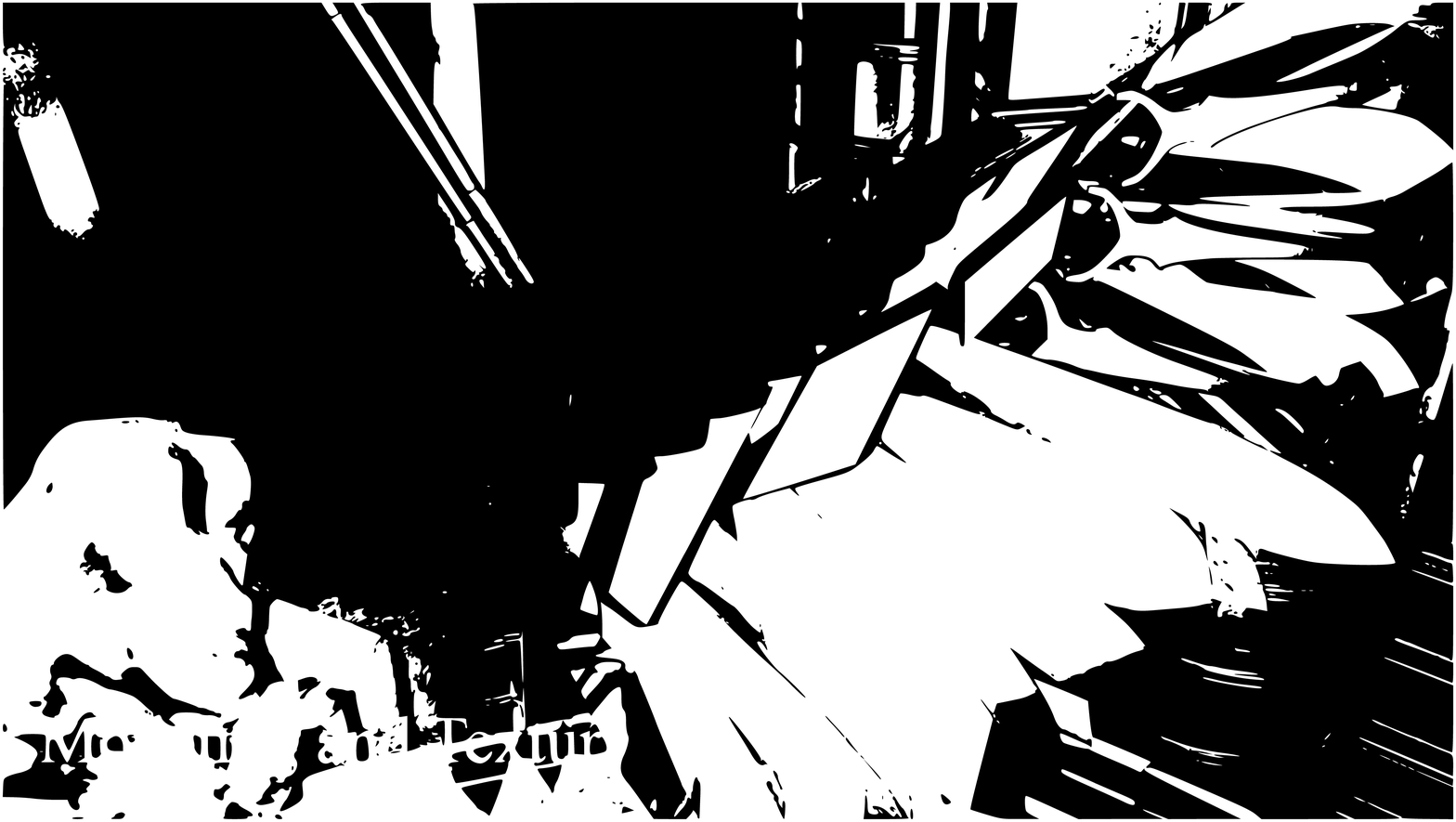}
    \includegraphics[width=0.3 \columnwidth]{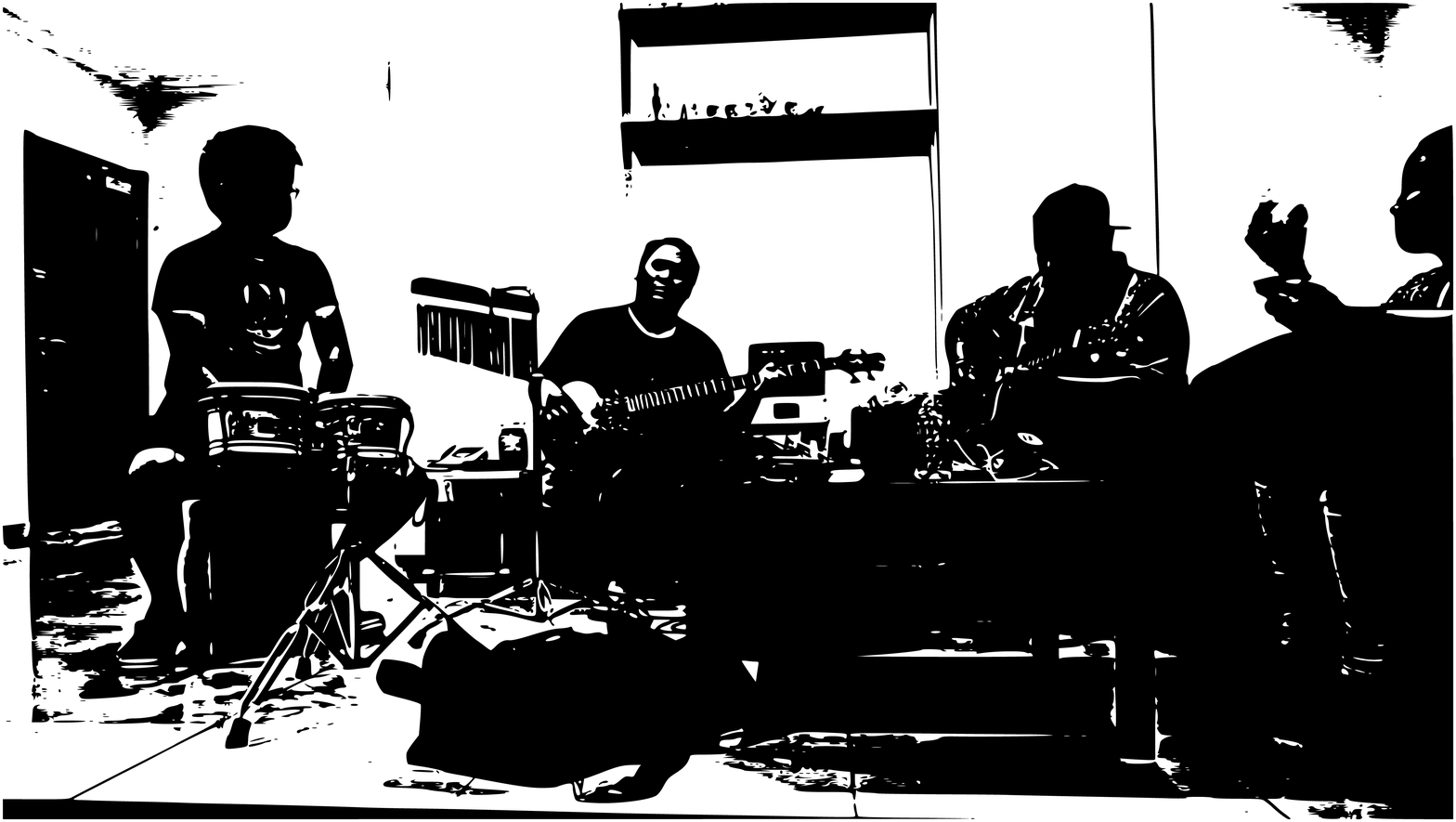}
    \includegraphics[width=0.3 \columnwidth]{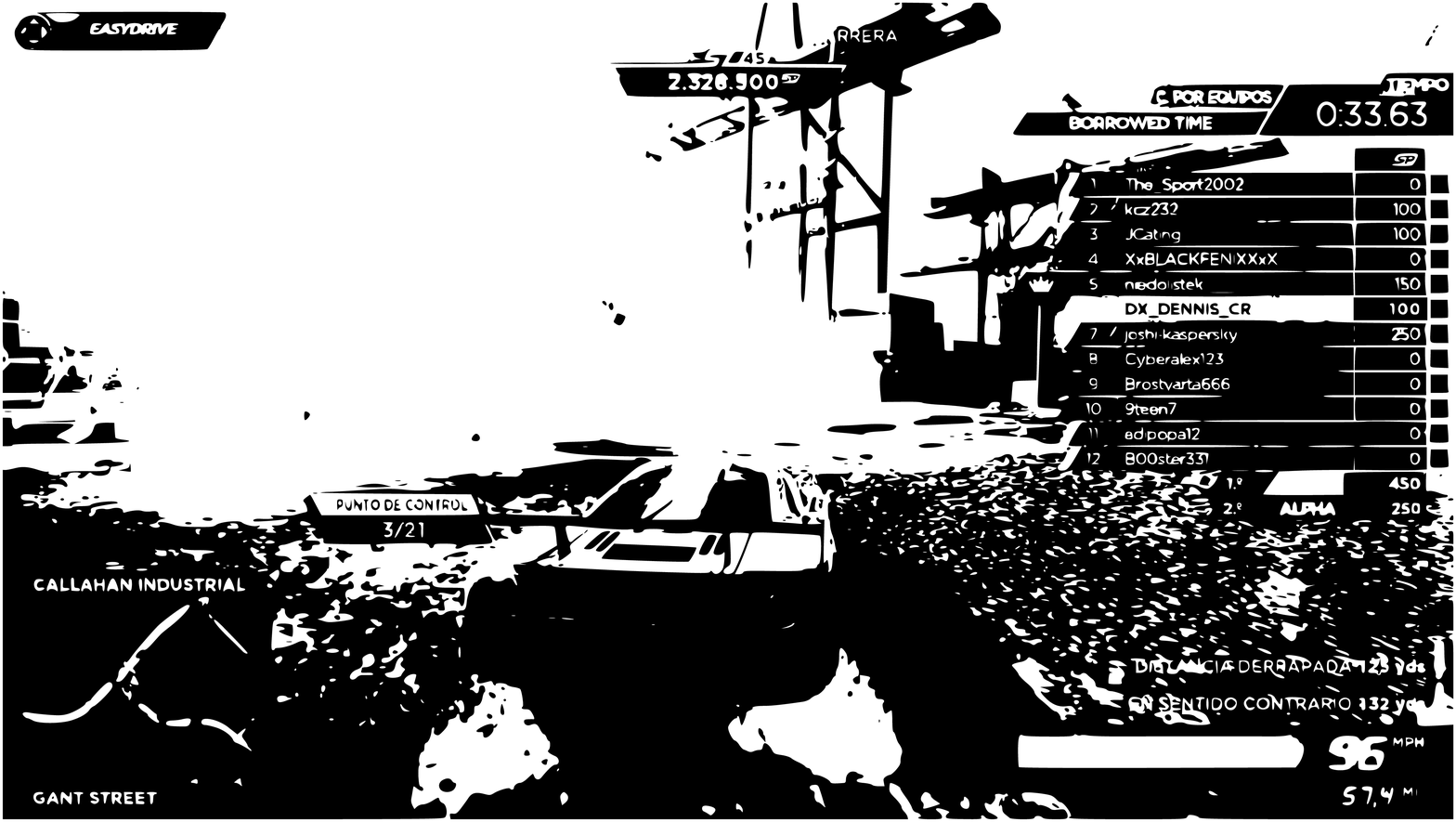}
     \\
     \includegraphics[width=0.3 \columnwidth]{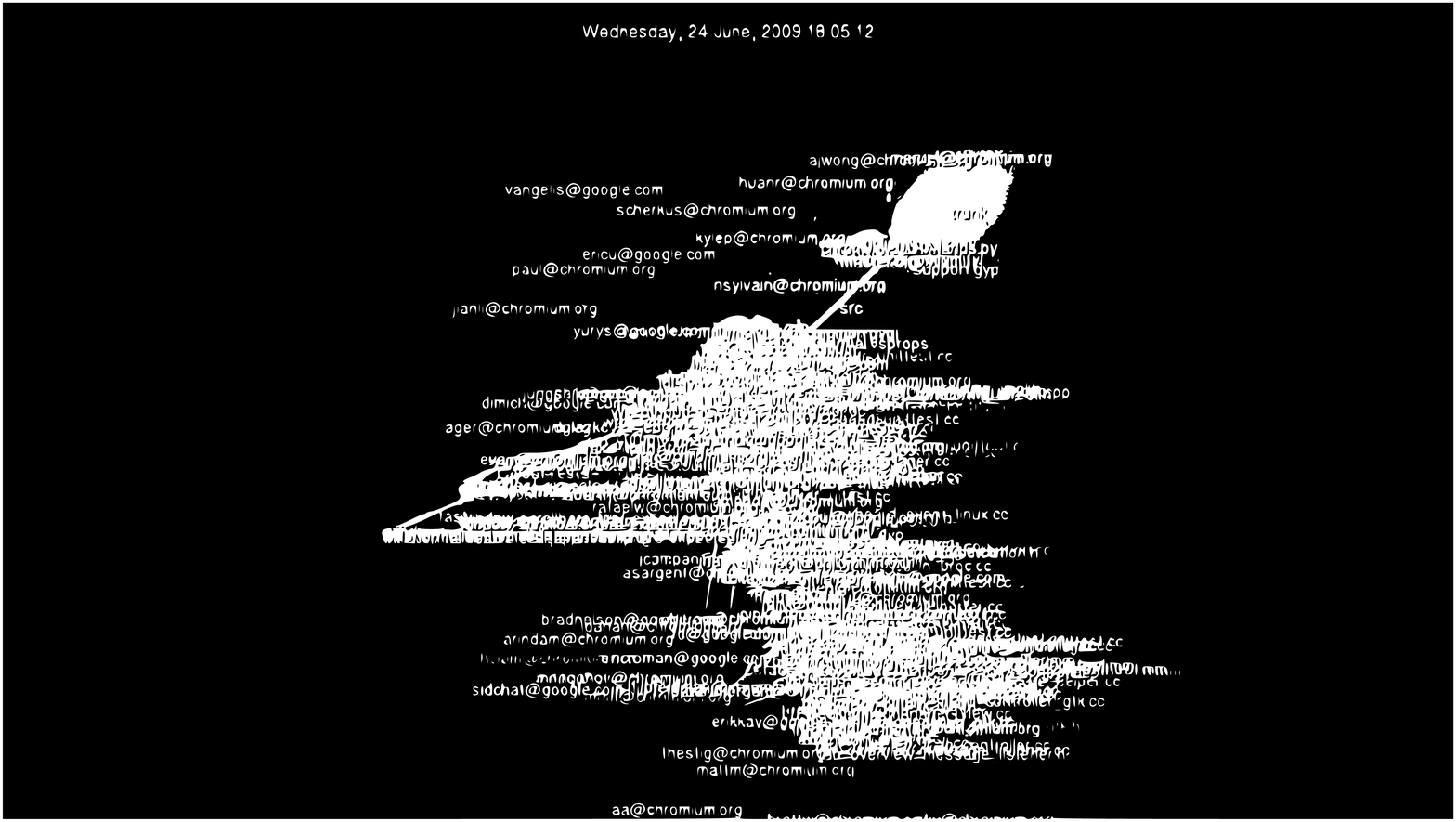}
    \includegraphics[width=0.3 \columnwidth]{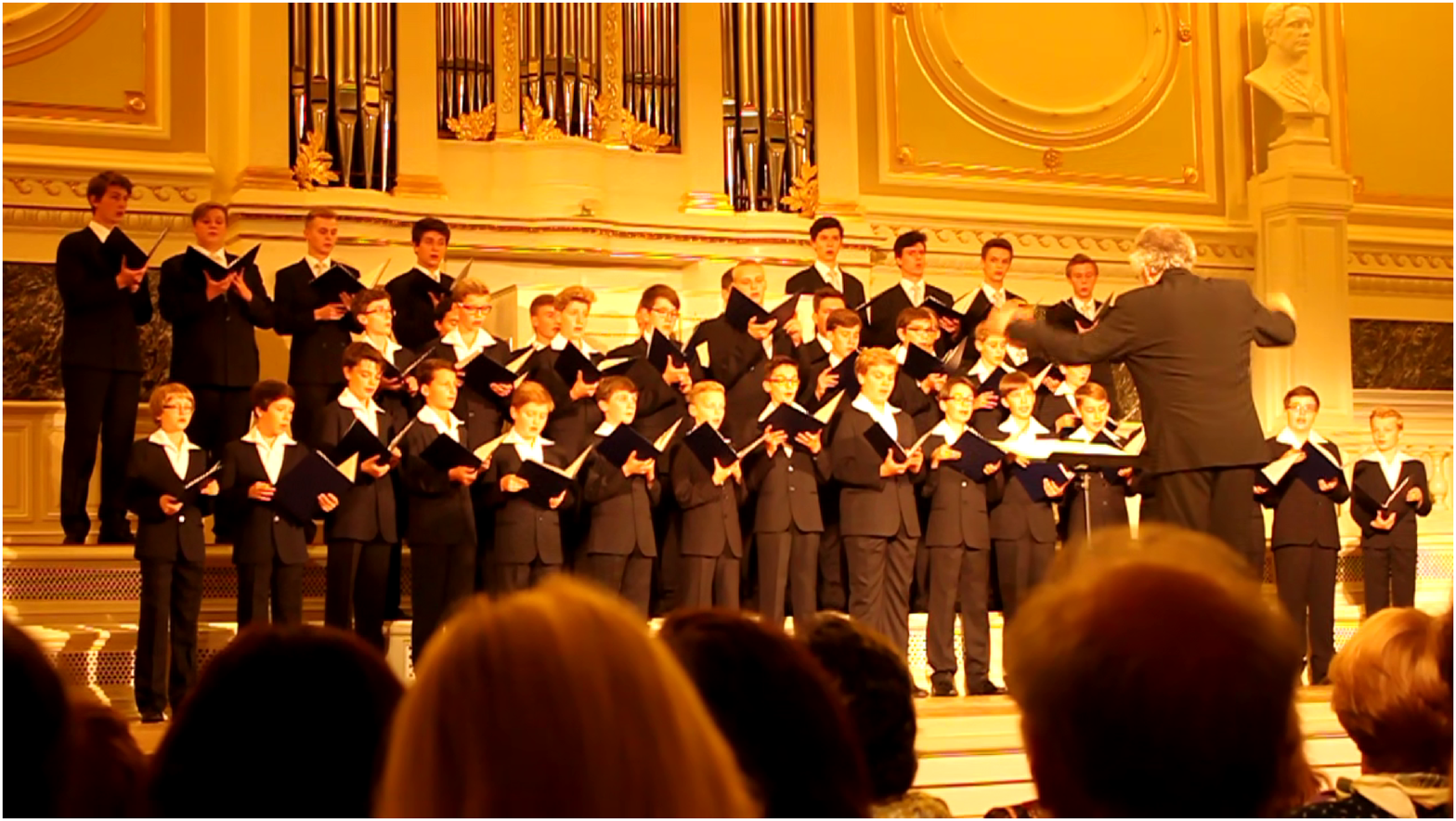}
        \includegraphics[width=0.3 \columnwidth]{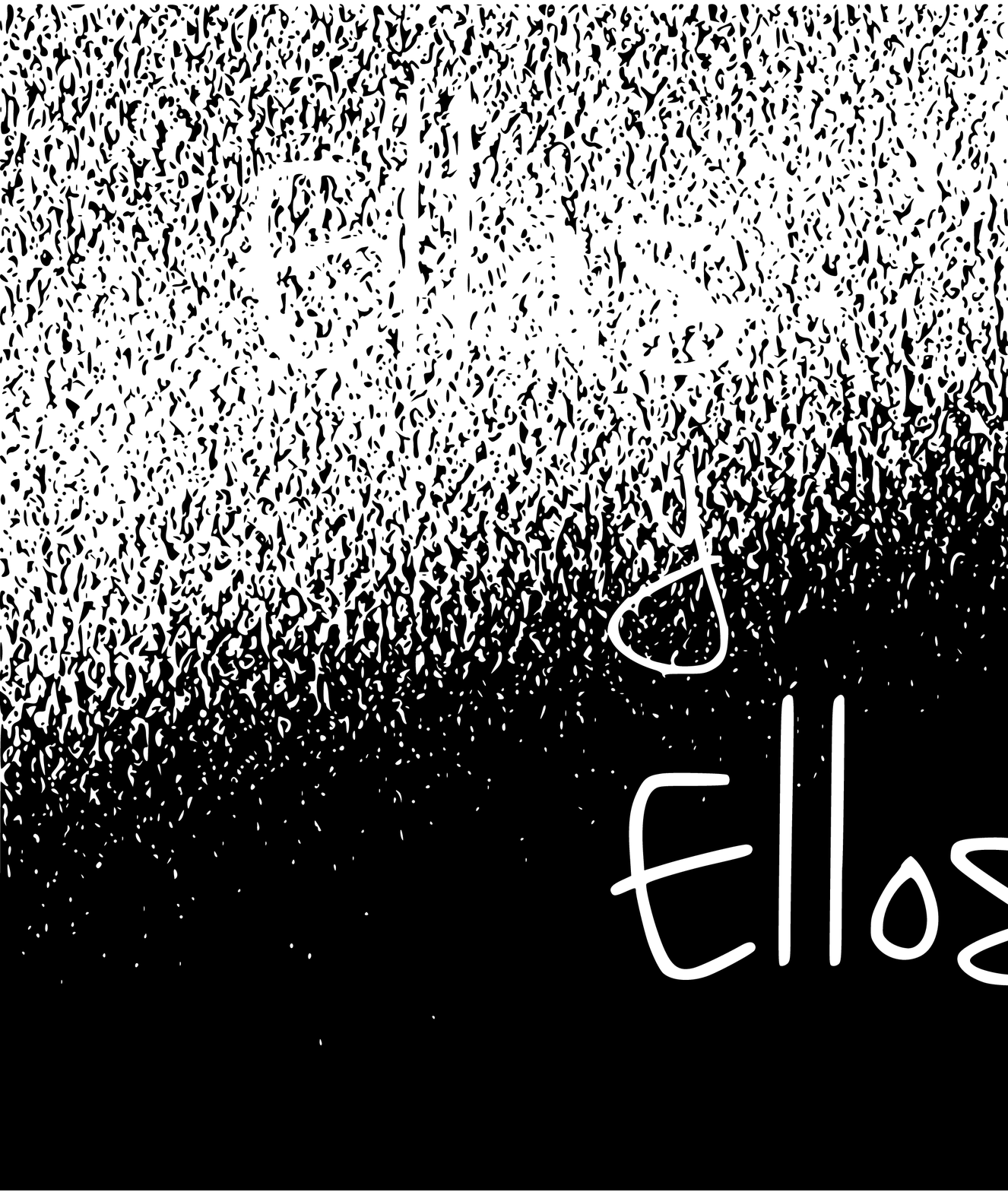}

    \caption{Example frames from different classes in the YouTube dataset. }
    \label{datasetExample}
\end{figure}

\section{Generating Ground Truth Optimal Lambda Values}

Using this direct optimisation technique, we  estimated in our previous work\cite{EIRingis}, the optimal value $k$ for a corpus of 77 clips for the x265 and showed that an improvement of 21\% could be obtained by fine tuning the Lagrangian Multipliers.
We propose here to extend the study to a much larger dataset of 10K clips from the YouTube UGC dataset,  covering thus a much wider range of use cases. We also  extend the study by considering both x265 and VP9.

\subsection{Dataset}

Previous work used a small corpus size (approximately 40 clips\cite{ma2016adaptive}\cite{zhang_bull} (up to 300 frames per clip)) as well as 77 clips in our own previous work\cite{EIRingis}. Also the types of content used in previous corpora are not necessarily a good representation of modern material. Therefore, in this work, we use an expanded dataset of 9,476 5-second clips.
This includes the recently published YouTube dataset \cite{wang2019youtube} representing 12 classes of video as specified by the YouTube team. A sample of each class of the clips can be seen in Figure \ref{datasetExample}. In addition we use clips from other publicly available datasets, these include Netflix dataset (Chimera and El Fuente)\cite{netflixdb}, DynTex dataset\cite{dyntex}, MCL\cite{MCL} and Derfs dataset\cite{derf}. Multiple DASH segments (clips) of 5 seconds (150 frames) were created from each sequence. Table \ref{clipbreakdown} shows the database composition by clip categories. There are 9,746 video clips at varying resolutions with a wide range of video content, representative of typical usage. This represents more than a 100 fold increase in the amount of data used for our experiments.

\begin{table}[t]
\centering
\caption{Composition of the dataset. Each segment is 150 frames (5 seconds) chosen to represent a typical DASH streaming segment length. \label{clipbreakdown}}
\begin{tabular}{lll}
\toprule 
\textbf{Subset} &\textbf{Segments} & \textbf{Resolution} \\ \midrule
Youtube UGC \cite{wang2019youtube}     &   8000      &         320p to 1080p       \\ 
MCL \cite{MCL}   &    220     &         1080p            \\ 
Dyntex   \cite{dyntex}       &           650       &        240p         \\ 
Derfs   \cite{derf}     &          796      &         240p to 1080p     \\ 
Netflix  \cite{netflixdb}     &         80        &      1080p         \\ \midrule
\textbf{Total} & \textbf{9746} &\\\bottomrule
\end{tabular}
\end{table}

\subsection{Implementation}

The codebase for x265 and VP9 was modified to take \(k\) as an argument. This allows the default $\lambda_{orig}$ to be modified according to equation \ref{kfactor}. We use the VideoLAN \cite{videoLAN} implementation of H.265, {\tt x265}\footnote{Version: 3.0+28-gbc05b8a91} and the WebM Project \cite{vp9webm} implementation of {\tt VP9}\footnote{Version: v1.8.1-152-g65c439523}. It is important to note that this work does not aim to compare HEVC and VP9, but to improve both codecs relative to their default behaviour. The following command invocations were used:

\textbf{HEVC:} \texttt{x265 --input SEQ.y4m --crf <XX> --tune-PSNR --psnr --output OUT.mp4 }

\textbf{VP9:} \texttt{vpxenc -p 1 --end-usage=vbr --cq-level <XX> --tune=psnr --psnr -o OUT.mkv SEQ.y4m}

where {\tt SEQ} and  {\tt OUT}  are the filenames for the raw input file and output encoded video. Clips were encoded with {\tt <XX>}  in the range 22:5:42.

%In our prior work\cite{EIRingis}, we indicated that Brent's method was the best, for this section all reported results are using this direct optimisation algorithm. We examine best case results  illustrated in Figure \ref{CDFPlotD}.

\subsection{Results}

\begin{figure}
    \begin{tabular}{cc}
    \includegraphics[width=0.49\linewidth]{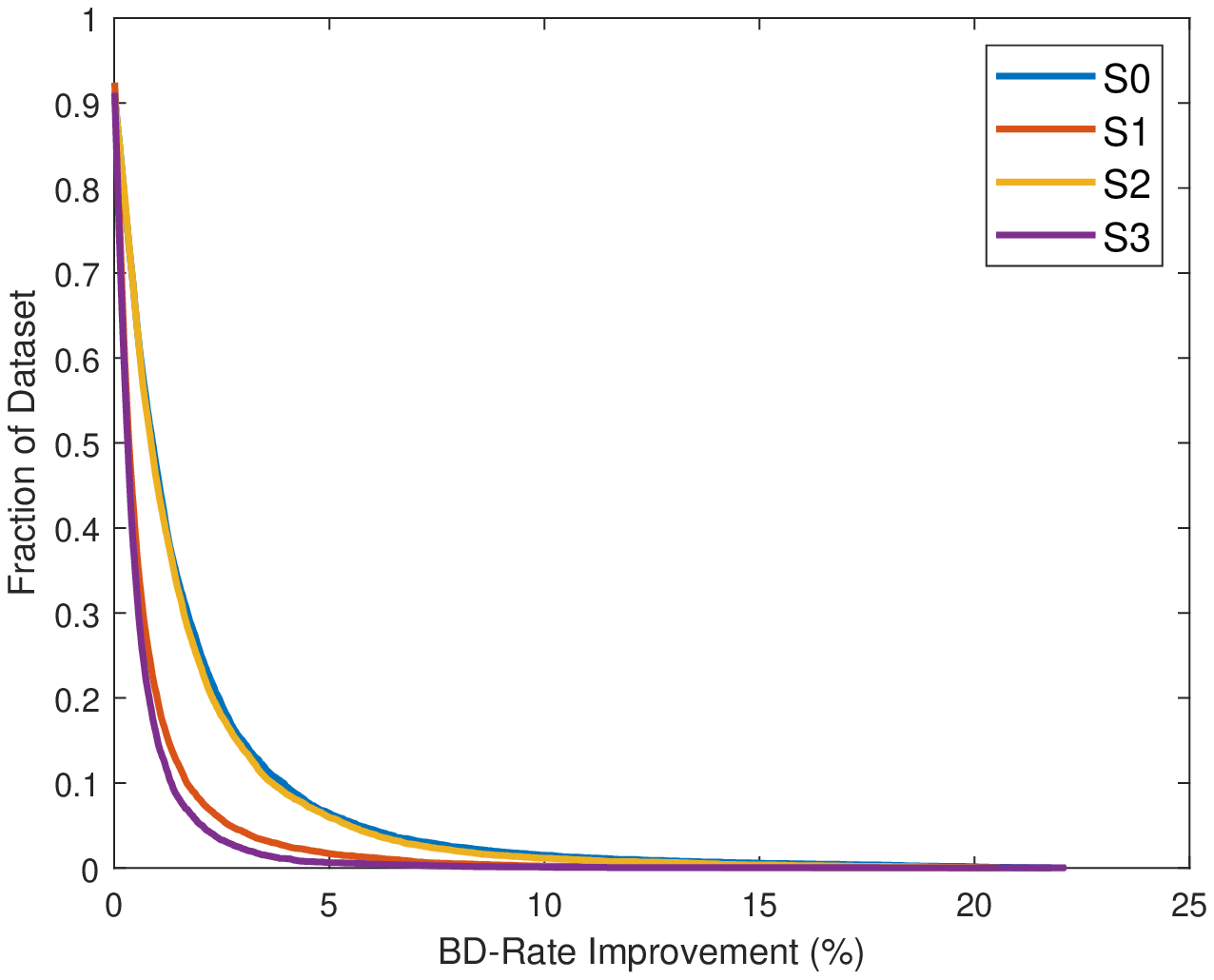}&
    \includegraphics[width=0.49\linewidth]{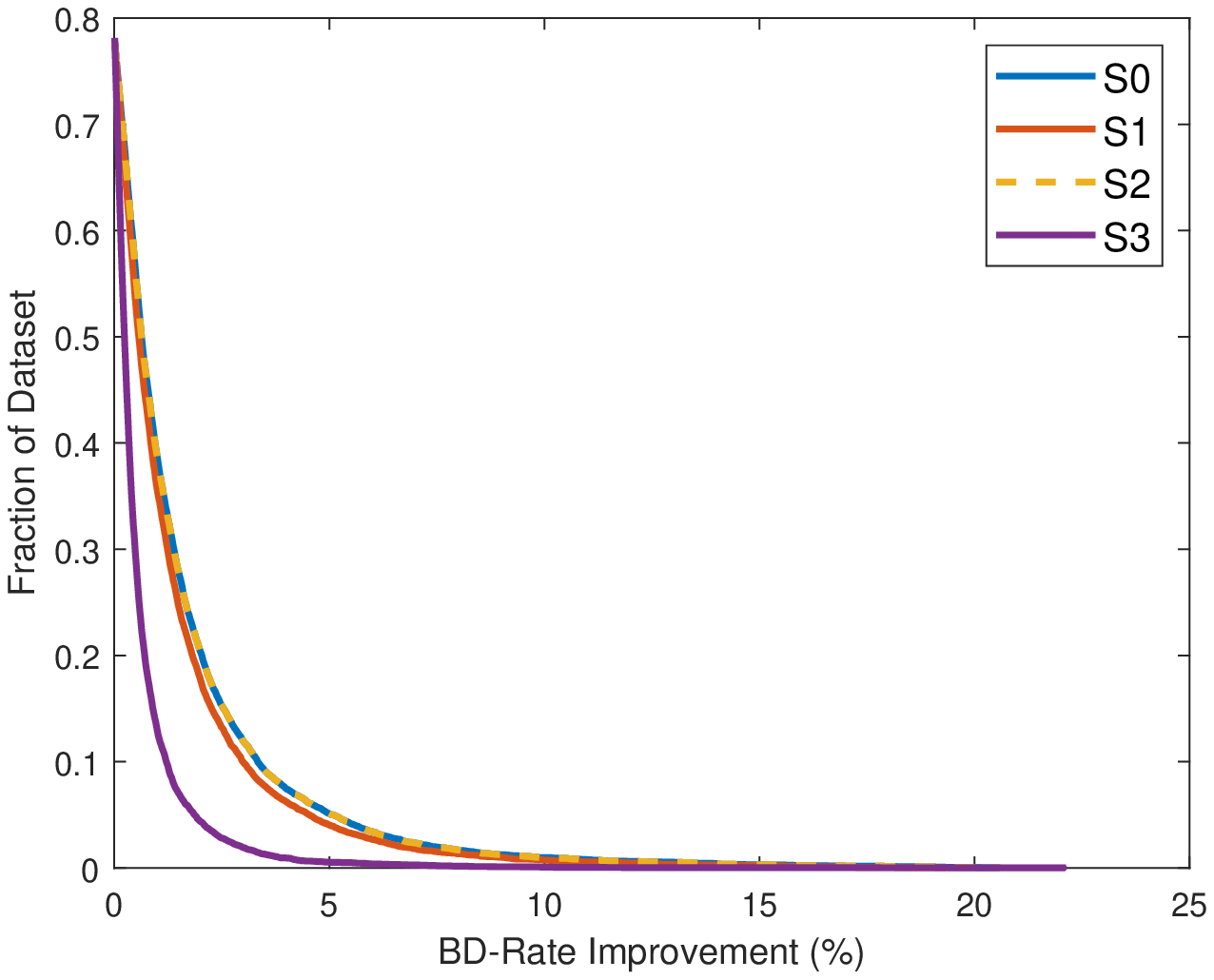}\\
            (a): HEVC & (b): VP9 
    \end{tabular}
    \caption{Cumulative distribution of BD-Rate improvement over 9,746 clips using $k$ calculated from various versions of the data or different codec invocations. Left: HEVC and right: VP9. The best performing systems would be nearest to the top right of the plot. In blue is the BD-Rate distribution using the original resolution and $k_D$ i.e. the ground truth best performance possible. Using faster video encodes but at the original resolution (system {\bf S2} yellow) yields almost identical performance to ground truth but with improvement in speed of 6\% for VP9 and 300\% for HEVC. The other curves show use of smaller resolutions and/or H264 as the codec. They are much faster (up to $\times 22$) but as can be seen achieve about 33\% of the possible gains on average. }
    %{\bf FP: I would remove the arrow in the graphs. This is quite distracting and the meaning of it is not immmediately obvious}}
    \label{CDFPlotD}
\end{figure}

For our corpus of video clips we minimise BD-Rate w.r.t. $k$ following the algorithm in the previous section, applied to the original resolution data (system \textbf{S0}). The cumulative distribution of the resulting maximum improvement (minimum BD-Rate using the optimal $k = k_D$) is shown in Figure~\ref{CDFPlotD}. That figure actually shows a summary of improvements using proxies as well but in this section we are only interested in the curve labelled {\em S0} in blue. As can be seen the distribution is long tailed implying that most clips show moderate improvement (1.87\%, 1.324\% (HEVC, VP9) on average) and in fact approximately 20\% of clips encoded using \(k_{D}\) had a BD-Rate improvement of 1\% or more. The best improvement for HEVC and VP9 was 23.41\% and 22.13\% respectively. Of course there are still a number of video clips where the original Lagrangian multiplier was the best. Using a BD-Rate improvement of 0.01\% as the minimum level of improvement, 93\% of clips encoded with HEVC and 75.6\% encoded with VP9 showed improvement. This is consistent with our previous work\cite{EIRingis} which had up to 21\% BD-Rate improvement for HEVC with 87\% of the clips encoded having an improvement.

%{\bf FP: add reference to previous work. How do these new results compare to your previous work. Any difference worth mentioning? XXXX  }

% \begin{figure}
%     \centering
%     \includegraphics{images/x265direct.eps}
%     \caption{BD-Rate improvement vs fraction of dataset for $k_D$ computed for the full dataset. It can be seen that approximately a fifth of each dataset having 1\% or better BD-Rate improvement. }
%     \label{CDFPlot}
% \end{figure}

 This direct optimiser ({\bf S0}) took on average 11.7 iterations, which would mean $11.7 \times 5 \approx 60$ video encodes. This is manageable at small resolutions, but is computationally intensive at modern resolutions (720p and higher) and {\em at scale}. The next sections introduce our approaches to address this problem.

\section{Improving computational cost}

In order to make direct optimisation more computationally efficient, we consider both using a proxy video encode and Machine Learning methods using low bandwidth video features. Using proxies can be either an encoding of the raw video at a smaller resolution or using faster codec settings or using an older/faster codec (eg H264). Each of these would represent a reduction in computational cost, even when using the same direct optimisation algorithm previously outlined. The invocations used in this work are as follows.

\noindent \textbf{HEVC:} 

\texttt{x265 --input  SEQ.y4m --tune psnr --psnr  --crf Q  --frames 150  --output  OUT.mp4}

\noindent\textbf{VP9:} 

\texttt{ vpxenc -p 1 --end-usage=cq --threads=7 --tune=psnr --psnr --cq-level=Q OUT.mkv SEQ.y4m}

\noindent\textbf{H264:}

\texttt{ x264 --frames 150 --tune psnr --psnr  --crf Q --output  OUT.264 SEQ.y4m } 

We present our new systems next.

\subsection{Using Proxies}
 We use clips at 144p as well as using the faster preset for the codecs (``ultrafast'' (H265) or ``rt''(VP9)). We also consider using H264 at original resolution as an alternate but faster codec. We organise the use of these proxies in the following systems. 
\begin{enumerate}
     \item[\textbf{S0}:] \(k_\textrm{D}\) estimated at original resolution/codec provides the reference best performance i.e. ground truth. This was presented in the previous section.
     \item[\textbf{S1}:] \(k_\textrm{144}\) estimated at 144p and used for encoding at the original resolution
     \item[\textbf{S2}:] \(k_\textrm{Dfast}\) estimated using the ``ultrafast'' (H265) or ``rt''(VP9) setting in the codec and used for encoding at the original resolution 
     \item[\textbf{S3}:] \(k_\textrm{264}\) estimated using H264 and used for encoding in the original codec
 \end{enumerate}
We implement the direct optimisation algorithm as in Section~\ref{dopt} for our three proxy systems (S1,S2,S3). FFMPEG is deployed with bicubic interpolation for downsampling the video corpus to 144p resolution for S1.  We use the ``ultrafast'' preset for x265 and VP9 for S2, and we modify x264\footnote{Version: 0.157.x} in a similar fashion to x265 by inserting our own multiplier $k$.

\subsubsection{Speed Gain with Proxies}
In order to quantify the savings in compute time achieved by using a proxy we encoded 250 DASH clips (150 frames) of the sequence Elephants Dream at 720p with each proxy. This time was measured on the same machine (Intel(R) Xeon(R) CPU E3-1240 v5 @ 3.50GHz
with 16GB RAM) and the relative performance of a single iteration in each system is shown in Table~\ref{compare}. 
% \begin{table}[t]
% \centering
% \caption{Processing time comparison of VP9 and H265 with faster codec settings and smaller resolutions as well as H264 \label{compare}} \vspace{.5em}
% \begin{tabular}{llllllll}
% \toprule
% \textbf{HEVC} & \textbf{Time (s)} & \textbf{per clip} & \textbf{Speedup} & \textbf{VP9} & \textbf{Time (s)} & \textbf{per clip} & \textbf{Speedup} \\ \midrule
% \textbf{S0} & 4871 & 19.48 & 1 & \textbf{S0} & 12573 & 50.29 & 1    \\ 
% \textbf{S1} & 222 & 0.89 & 21.94 & \textbf{S1} & 1406 & 5.62 & 8.94 \\ 
% \textbf{S2} & 1585 & 6.34 & 3.07 & \textbf{S2} & 11796 & 47.18 & 1.07 \\ 
% \textbf{S3} & 1348 & 5.39 & 3.61 & \textbf{S3} & 1348 & 5.39 & 9.33       \\ \bottomrule
% \end{tabular}
% \end{table}
\begin{table}[t]
\centering
\caption{Processing time comparison of VP9 and H265 with faster codec settings and smaller resolutions as well as H264 \label{compare}} \vspace{.5em}
\begin{tabular}{lrrr}
\toprule
\textbf{System} & \textbf{Time (s)} & \textbf{per clip} & \textbf{Speedup} \\ \midrule
\textbf{HEVC: S0} & 4871 & 19.48 & 1 \\
\rowcolor{Gray}\textbf{HEVC: S1} & 222 & 0.89 & 21.94  \\ 
\textbf{HEVC: S2} & 1585 & 6.34 & 3.07 \\ 
\rowcolor{Gray}\textbf{HEVC: S3} & 1348 & 5.39 & 3.61  \\ \midrule
\textbf{VP9: S0} & 12573 & 50.29 & 1    \\ 
\rowcolor{Gray}\textbf{VP9: S1} & 1406 & 5.62 & 8.94 \\
\textbf{VP9: S2} & 11796 & 47.18 & 1.07 \\
\rowcolor{Gray}\textbf{VP9: S3} & 1348 & 5.39 & 9.33  \\
\bottomrule
\end{tabular}
\end{table}

From table \ref{compare}, we can see that using the smaller resolution videos gives us the biggest improvement in speed with the time to encode being 22x faster with x265 and 9 times faster with VP9. Encoding using x264 also shows a significant improvement in encode time with 3.7x and 9.3x (HEVC, VP9) times improvement. The smallest savings are found when using the faster encoding presets for each codec with 3.1x and 1.1x (HEVC, VP9) improvements being found using the ``ultrafast'' and ``rt'' presets respectively. If we are able to get comparable BD-Rate savings using these proxies it would indicate a significant improvement in compute time for the direct optimisation methods applied. 

\subsubsection{Performance Gains with Proxies}

\begin{figure}
    \centering
    \includegraphics[width=0.45\columnwidth]{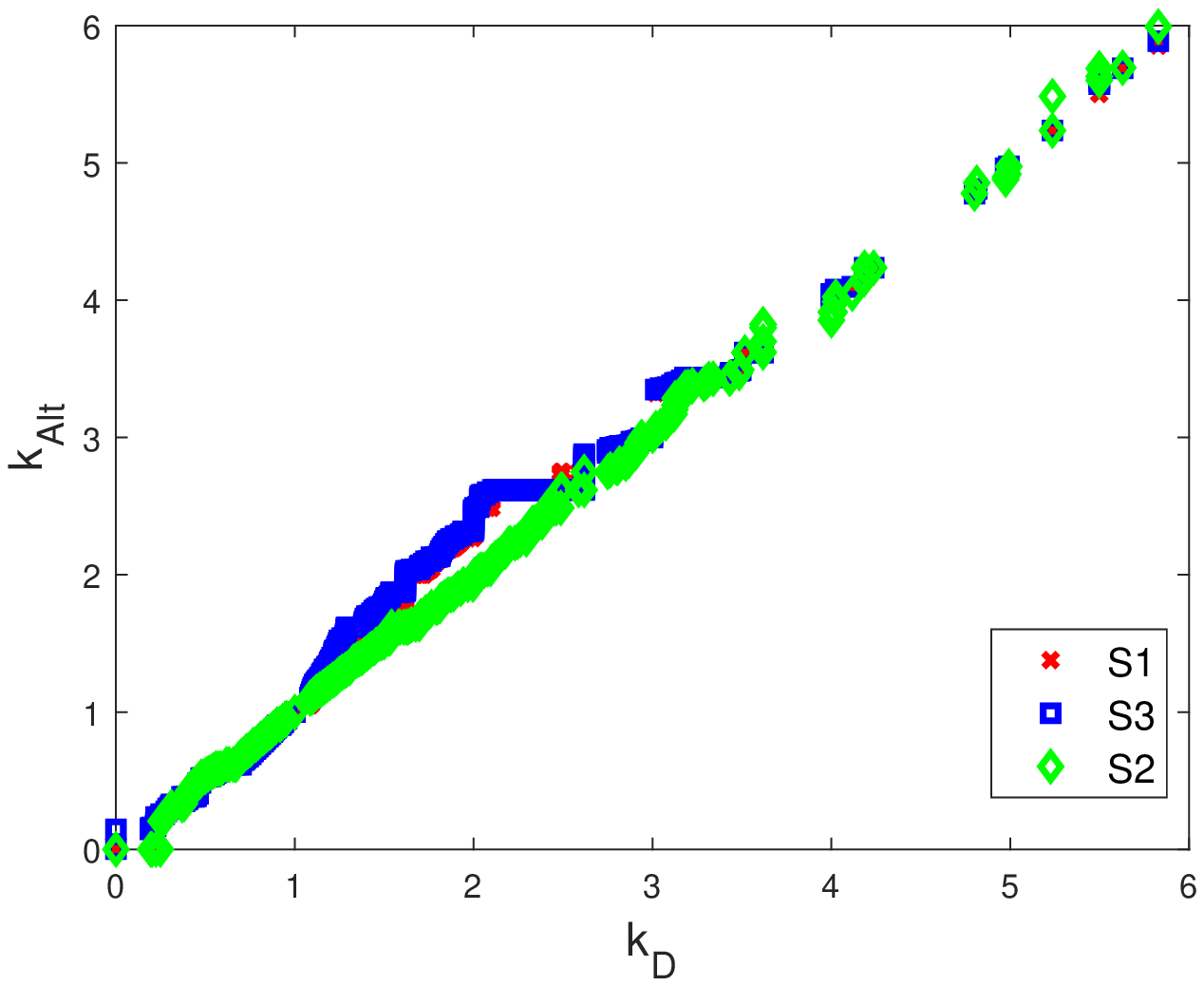}
    \includegraphics[width=0.45\columnwidth]{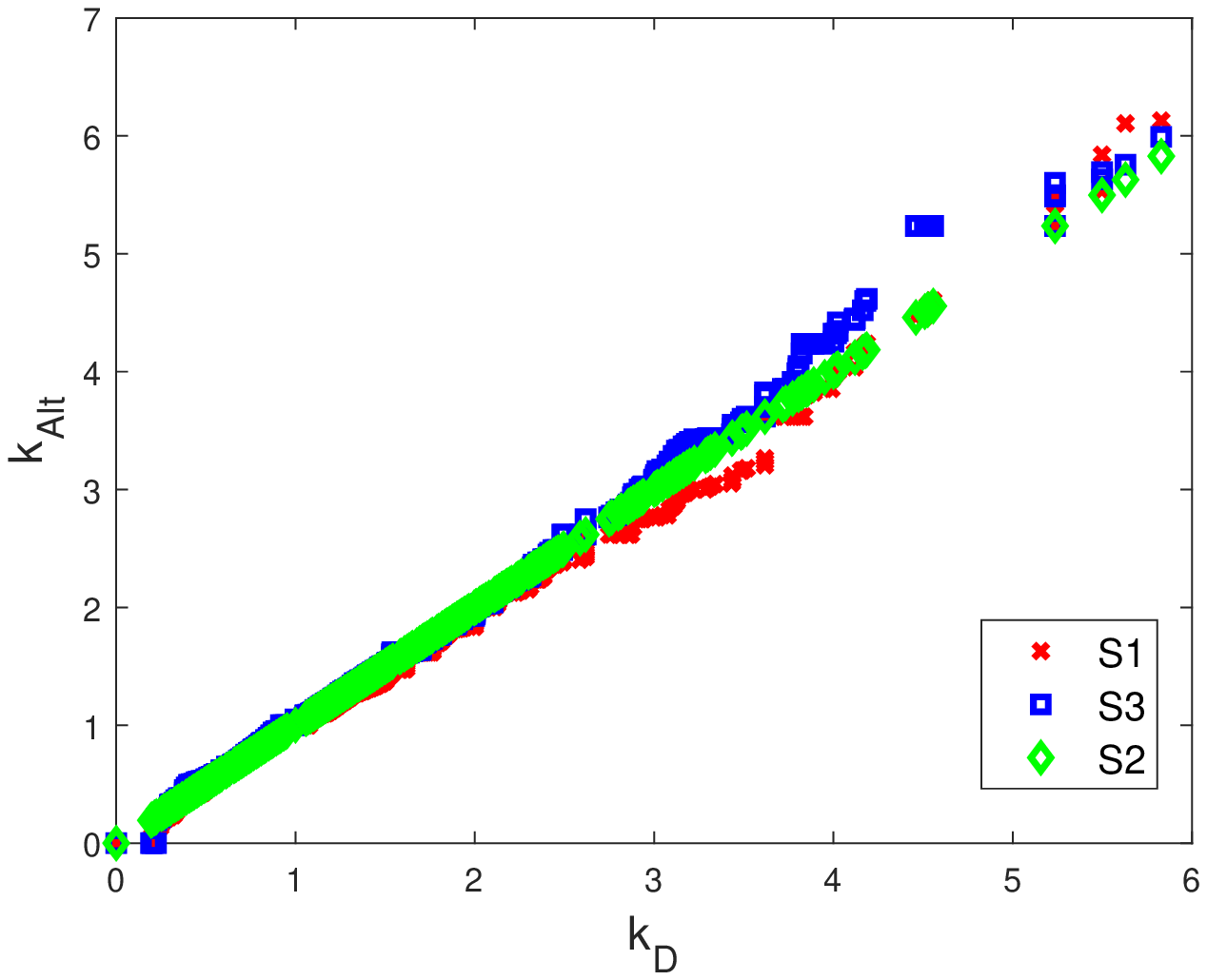}
    \caption{ \label{kCompareProxy} $k_D$ vs $k_{\tt Alt}$ determined from a proxy video or codec LEFT: x265, RIGHT: VP9. Ideally, we would want points to lie on the line y=x. We can see a strong correlation between $k$ and $k_{\tt Alt}$ determined by \textbf{S2} (green) and \textbf{S0}.}% . {\bf XXX FP: is green S0? I thought $k_D$ is S0 and green is S2?? XXXX. By the way, why using $k_D,k_{144},etc.$ and not $k_0,k_1,etc.$?} }
   
\end{figure}

Figure \ref{kCompareProxy} shows that for the majority of clips there is a close correlation between ground truth $k=k_D$ generated at original resolution and $k$ generated at 144p. This is because the scatter plots are close to the $k=k_D$. This holds true for all three systems S1, S2 and S3 for both HEVC and VP9. $k$ is most similar to the original estimates when using the faster codec settings at the original resolution (S2) and least similar using H264 to estimate (S3). 

Figure \ref{CDFPlotD} summarises the BD-Rate performance gains over the whole corpus. The Cumulative density distribution of each system is reported there. Using the fast presets at the original resolution (S2) gives us very similar performance to our original direct optimisation algorithms in both HEVC and VP9. Using a different codec (H264) to estimate $k$ however, gave the worst results. A summary of our findings using systems S1, S2, S3 can be found in table \ref{DirectSummary}.

\begin{table}[t]
\centering

\caption{\label{DirectSummary} Summary of {\bf BD-R Gains} results using Direct Optimisation with Proxies. This shows the $\%$ of clips which have no improvement, $\> 0.1\%$ and $\>1\%$ improvement, as well as the best and average BD Rate improvement. * indicates which system we believe provides us with the best balance between speedup and maintaining BD-Rate improvement  }
\vspace{.5em}
\begin{tabular}{lrrrrrr}
\toprule
\multirow{2}{*}{\textbf{System}} & \multicolumn{3}{c}{\bf Clips with BD-R gain of} & {\textbf{Best}} & {\textbf{Avg}} & \multirow{2}{*}{\textbf{Speedup}} \\
 {} & \textbf{$=0\%$} & \textbf{$>0.1\%$} & \textbf{$>1\%$} &
{\bf Gain} & {\bf Gain} & {}\\ \midrule
\textbf{S0: x265}  & 8\% & 87.00\%  & 46.24\% & 23.86\%  & 1.87\% & $\times$1.0\\ 
\rowcolor{Gray}\textbf{S1: x265} & 9\% & 78.17\% & 45.23\% & 20.46\% & 0.54\% & $\times$21.9 \\ 
\textbf{S2: x265 *} & 9\% & 86.57\% & 18.75\% & 20.59\% & 1.54\%  & $\times$3.1 \\ 
\rowcolor{Gray}\textbf{S3: x265}  & 9\% & 74.21\% & 14.13\% & 20.12\% & 0.58\% & $\times$3.6    \\\midrule
\textbf{S0: VP9} & 22\%   & 74.00\%   & 38.75\%   & 22.13\%     & 1.63\%  & $\times$1\\ 
\rowcolor{Gray}\textbf{S1: VP9}  & 22\%    & 73.00\%   & 35.73\%     & 20.60\%    & 0.5\%   & $\times$8.9 \\ 
\textbf{S2: VP9}   & 22\%    & 74.00\%   & 38.75\%   & 22.13\%     & 1.32\% & $\times$1.1    \\ 
\rowcolor{Gray}\textbf{S3: VP9 *}   & 22\% & 66.00\%    & 12.85\%  & 19.36\%    & 1.19\%  & $\times$9.3 \\ \bottomrule
\end{tabular}
\end{table}

In all the proposed systems, the majority of clips show some BD-Rate improvement. The average BD-Rate improvement is in the range of 1\% across the corpus. We see in Table \ref{DirectSummary} that the best tradeoff is system S3 for VP9 (1.19\% BD-Rate avg gain out of a possible 1.63\% with S0, thus 73\% of the potential average BD-Rate gains at $\times$ 9.3 speed) and S2 for HEVC (83\% of the possible average BD-Rate gains from S0 at $\times$ 3.1 speed). 

Using the direct optimisation methods still requires a large number of video encodes for a given clip regardless of resolution. Hence the next level of performance is to use a Machine Learning approach.

\subsection{Machine Learning}
We investigate an Offline process in which the first step measures video features that are  mapped to $k$ through an ML algorithm. This significantly reduces computational complexity. %In essence we are generalising the previous approaches to reach beyond regression in estimation of $k$.
Our proposed feature set consists of 17 features: Overall Bitrate, Overall PSNR, PSNR for Y, U and V channels, Avg bitrate for I, P and B frames, Avg PSNR (Y,U,V) for I, P and B frames, Frame height and width.
Similar to Covell et al \cite{covell2016optimizing}, non linear extensions of the feature set were created from products of feature elements by taking the product of a subset of feature pairs. The final feature vector therefore contains 49 elements. 

We use \(k_D\) as ground truth for training our ML process. Note that $k$ was range limited  $0<k<6$ and we hold out 10\% of the dataset for testing. With the remaining corpus, cross validation with five (5) folds was done for the training to minimize over-fitting. We tested 19 ML methods (including Logistic and Linear Regression SVMs, Decision Trees) and found that Random Forests yielded best results. We can denote these additional systems created using machine learning results as follows:

\begin{enumerate}
     \item[\textbf{ML0}:] \(k_\textrm{M}\) using features estimated at original resolution/codec 
     \item[\textbf{ML1}:] \(k_\textrm{M144}\) using features estimated at 144p and used for encoding at the original resolution
     \item[\textbf{ML2}:] \(k_\textrm{Mfast}\) using features estimated using the ``ultrafast'' (H265) or ``rt''(VP9) setting in the codec and used for encoding at the original resolution 
 \end{enumerate}

Using ML0, the random forest model as stated above, we find that 64.74\%, 62.38 \% (HEVC, VP9) of the possible BD-Rate gain at original resolution (i.e. S0) was achieved. Furthermore, 66.96\%, 58.22 \% (HEVC, VP9) of the video corpus showed an improved BD-Rate. Similar results were found with ML2, which used the results from the ''ultrafast`` encodes with 63.48\%, 57.79 \% (HEVC, VP9) of the possible BD-Rate gain at S0 were achieved. 

The biggest savings in computational time would be ML1, where we extract features at 144p, and use these to estimate $k$, ($k_\textrm{Msmall}$), and encode at the original. We find that  56.82\%, 54.27\% (HEVC, VP9) of the video corpus showed a BD-Rate improvement. Overall this implied that 64.24\%, 67.72\% (HEVC, VP9) of the possible gain was achieved.

Table \ref{MLSummary} shows a summary of the findings for systems ML1, ML2 and ML3. With nearly half of clips showing no improvement, we can see that this needs more work. Determining what features of the video are best suited to predict the appropriate $k$ for RD optimisation is a challenge. However, this work shows promise by being sixty (60) times faster than the direct optimisation methods in the previous section as only a single video encode is required to predict $k$ for a given clip. 

%\textbf{XXXX INSERT ML RESULTS IN TABLE LIKE TABLE 3 AND ALSO GRAPH LIKE FIGURE 4. REMEMBER TO PUT THE CAPTION IN THE NEW FIGURE THAT IS MEANINGFUL XXX REMEMBER TO MAKE THE NOTE THAT THE ML SPEEDUP IS BECAUSE YOU ONLY NEED ONE ENCODE XXX}

%Using faster video encodes but at the original resolution (system {\bf S2} yellow) yields almost identical performance to ground truth but with improvement in speed of 6\% for VP9 and 300\% for HEVC. The other curves show use of smaller resolutions and/or H264 as the codec. They are much faster (up to $\times 22$) but as can be seen achieve about 33\% of the possible gains on average. 

\begin{table}[t]
\centering

\caption{\label{MLSummary} Summary of {\bf BD-R Gains} Results using Machine Learning with Proxies. This shows the $\%$ of clips which have no improvement(including those which performed worse), $\> 0.1\%$ and $\>1\%$ improvement, as well as the best and average BD Rate improvement  }
\begin{tabular}{lrrrrrr}
\toprule
\multirow{2}{*}{\textbf{System}} & \multicolumn{3}{c}{\bf Clips with BD-R gain of} & {\textbf{Best}} & {\textbf{Avg}}  & \multirow{2}{*}{\textbf{Speedup}}  \\
 {} & \textbf{$=0\%$} & \textbf{$>0.1\%$} & \textbf{$>1\%$} &
{\bf Gain} & {\bf Gain} & {} \\ \midrule
\textbf{ML0: x265}  & 33\% & 27.5\%   & 2.2\% & 8.54\%  & 0.19\%& $\times$60 \\ 
\rowcolor{Gray}\textbf{ML1: x265} & 43\% & 26.8\% & 2.1\% & 11.83\% & 0.18\%  & $\times$1314 \\ 
\textbf{ML2: x265 } & 37\% &26.6\% &2.4\%  & 13.93\% &0.19\%   & $\times$186 \\ 
\midrule
\rowcolor{Gray}\textbf{ML0: VP9}  & 41\%    & 44.1\%  & 3.1\%     &    9.62\%&    0.14\% & $\times$60\\ 
\textbf{ML1: VP9}   & 45\%    &  39.7\% &  3.2\%  & 10.09\%      & 0.14\%    &$\times$534 \\ 
\rowcolor{Gray}\textbf{ML2: VP9 }   & 42\% &   40.1\%  & 3.3\% & 8.41\%   & 0.14\% &$\times$66   \\ \bottomrule
\end{tabular}
\end{table}

\begin{figure}
    \begin{tabular}{cc}
    \includegraphics[width=0.49\linewidth]{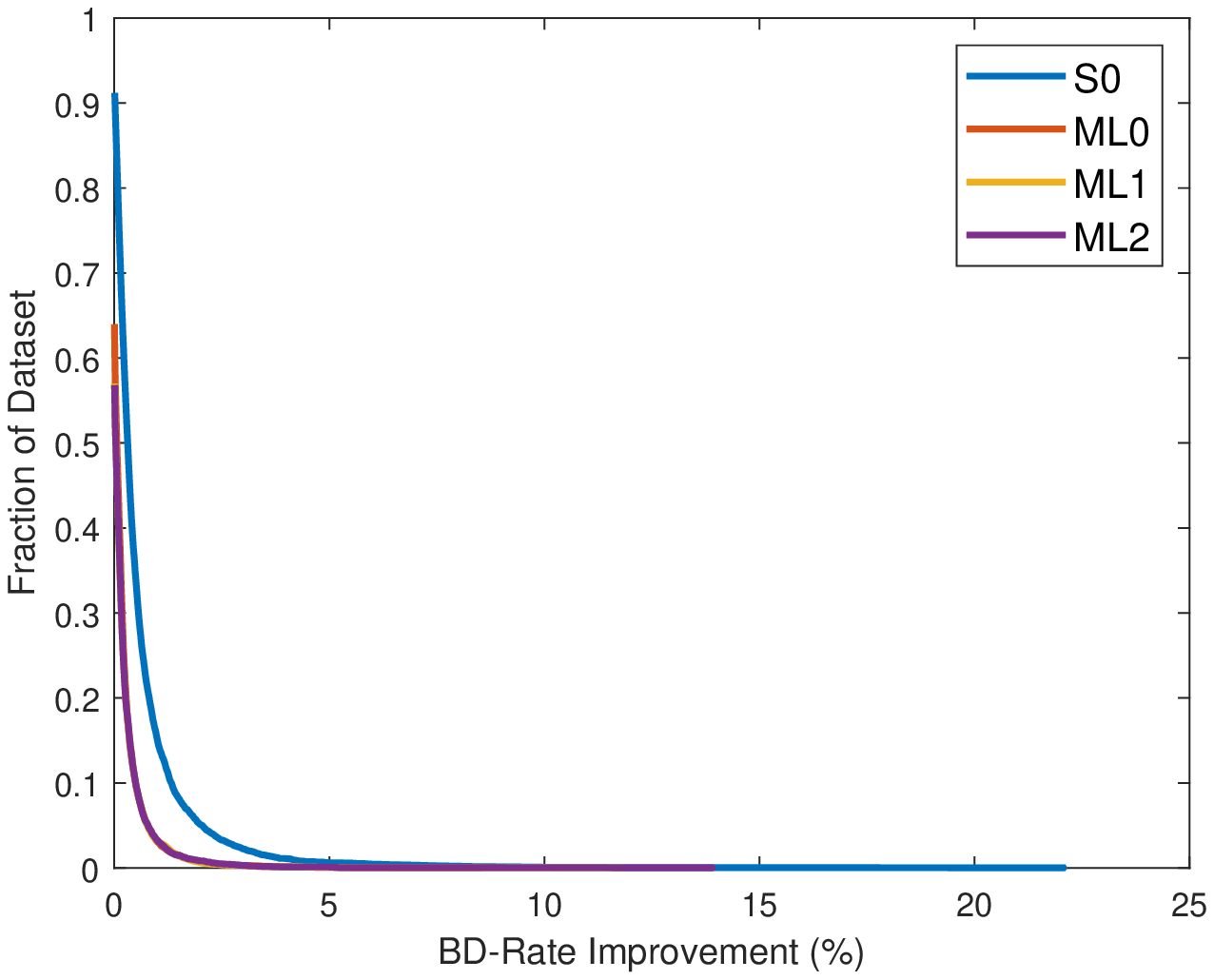}&
    \includegraphics[width=0.49\linewidth]{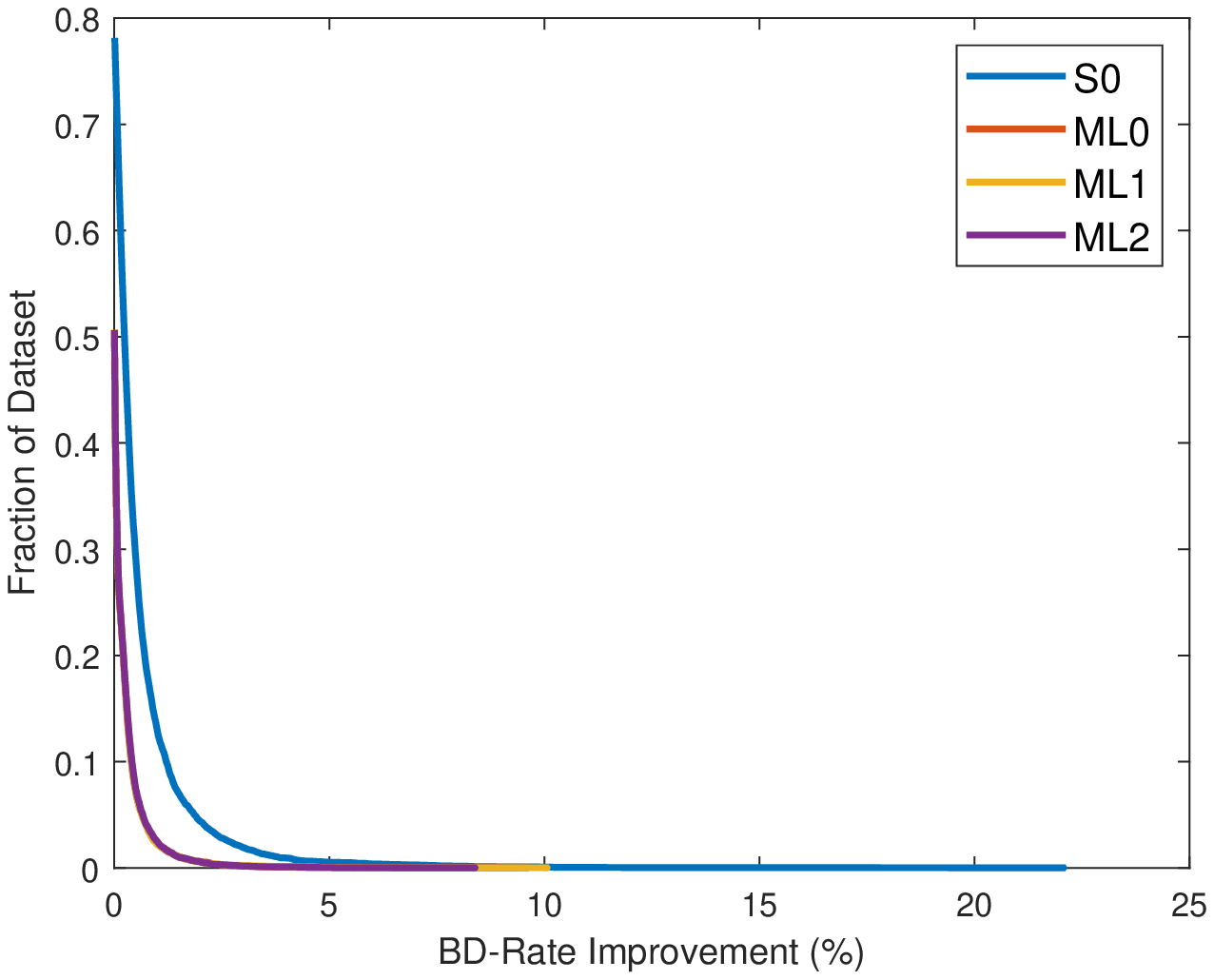}\\
            (a): HEVC & (b): VP9 
    \end{tabular}
    \caption{Cumulative distribution of BD-Rate improvement over 9,746 clips using $k$ calculated using the various Machine Learning systems. Left: HEVC and right: VP9. The best performing systems would be nearest to the top right of the plot. In blue is the BD-Rate distribution using the original resolution and $k_D$ i.e. the ground truth best performance possible. Over a third of the clips do not show any BD-Rate improvement. The three ML systems did not provide comparable gains to the direct optimisation scheme (S0), although they have an order of magnitude lower computational cost.}
    \label{CDFPlotM}
\end{figure}

\section{Conclusion}

%\textbf{XXX THIS BIT SEEMS LIKE BITS SHOULD BE IN A CONCLUSIONS SOMEWHEREXXXX}
%Overall, this work shows that BD-Rate gains are possible in many videos when the rate distortion equation is tailored on a per clip basis. We find that using proxies is useful and in fact can achieve over 60\% of the available gains. As can be seen up to 23\% improvement is possible (S0) and in most of these cases over 85\% of the possible gain is achieved by the most efficient system ML1. Given that the maximum gains are appreciable in all the systems, it seems that ML1 which is most efficient, is also a good compromise wrt BD-Rate gains. That is because instead of using about 60 encodes at the original videos, we use one encode at 144p and a Random Forest instantiation. 

In this paper, we have presented seven computationally efficient systems which estimate optimal $k$ in a variety of ways which rely on proxies and ML for estimation. It is important to note that this work does not aim to compare HEVC and VP9, but to improve both codecs using the same methods, with respect to their default configurations. Our best performing system wrt BD-Rate gain alone is direct optimisation (S0) at the target resolution. Gains of up to 23\% can be achieved here. That was used as the ground truth or upper limit of performance against which six more efficient systems were evaluated. We find that using proxies is useful and in fact can achieve comparable results at much faster speeds. It seems that S2 provides the closest estimate, but ML1 gives us the biggest improvement in compute time. Finally, we note that our corpus is more appropriate for modern use cases. The size of the data tested is significantly larger than previous work in this area. Future work will involve the exploration of more complex inference engines like DNNs.

%\appendix    %>>>> this command starts appendixes
\acknowledgments % equivalent to \section*{ACKNOWLEDGMENTS}       
 
This work was supported in part by YouTube, Google and the Ussher Research Studentship from Trinity College.

\bibliography{main} % bibliography data in report.bib

\begin{thebibliography}{10}

\bibitem{wang2019youtube}
{Wang}, Y., {Inguva}, S., and {Adsumilli}, B., ``Youtube {UGC} dataset for
  video compression research,'' in [{\em 2019 IEEE 21st International Workshop
  on Multimedia Signal Processing (MMSP)}{\nolinebreak\hspace{0.1em}]},   1--5
  (Sep. 2019).

\bibitem{cisco}
Cass, S., ``The age of the zettabyte cisco: the future of internet traffic is
  video [dataflow],'' {\em IEEE Spectrum}~{\bf 51}(3),  68--68 (2014).

\bibitem{sullivan2012overview}
Sullivan, G.~J., Ohm, J.-R., Han, W.-J., and Wiegand, T., ``Overview of the
  high efficiency video coding ({HEVC}) standard,'' {\em IEEE Transactions on
  circuits and systems for video technology}~{\bf 22}(12),  1649--1668 (2012).

\bibitem{mukherjee2013latest}
Mukherjee, D., Bankoski, J., Grange, A., Han, J., Koleszar, J., Wilkins, P.,
  Xu, Y., and Bultje, R., ``The latest open-source video codec {VP9}-an
  overview and preliminary results,'' in [{\em 2013 Picture Coding Symposium
  (PCS)}{\nolinebreak\hspace{0.1em}]},   390--393, IEEE (2013).

\bibitem{zhang2019overview}
Zhang, T. and Mao, S., ``An overview of emerging video coding standards,'' {\em
  GetMobile: Mobile Computing and Communications}~{\bf 22}(4),  13--20 (2019).

\bibitem{sullivan1998rate}
Sullivan, G.~J. and Wiegand, T., ``Rate-distortion optimization for video
  compression,'' {\em IEEE signal processing magazine}~{\bf 15}(6),  74--90
  (1998).

\bibitem{wiegand2001lagrange}
Wiegand, T. and Girod, B., ``Lagrange multiplier selection in hybrid video
  coder control,'' in [{\em Proceedings 2001 International Conference on Image
  Processing}{\nolinebreak\hspace{0.1em}]},   {\bf 3},  542--545, IEEE (2001).

\bibitem{ortega1998rate}
Ortega, A. and Ramchandran, K., ``Rate-distortion methods for image and video
  compression,'' {\em IEEE Signal processing magazine}~{\bf 15}(6),  23--50
  (1998).

\bibitem{wiegand1996rate}
Wiegand, T., Lightstone, M., Mukherjee, D., Campbell, T.~G., and Mitra, S.~K.,
  ``Rate-distortion optimized mode selection for very low bit rate video coding
  and the emerging {H. 263} standard,'' {\em IEEE Transactions on Circuits and
  Systems for Video Technology}~{\bf 6}(2),  182--190 (1996).

\bibitem{vp9webm}
Project, T.~W., ``{The WebM Project, {VP9} Project Summary}.''
\newblock https://www.webmproject.org/vp9/.

\bibitem{ma2016adaptive}
Ma, C., Naser, K., Ricordel, V., Le~Callet, P., and Qing, C., ``An adaptive
  lagrange multiplier determination method for dynamic texture in {HEVC},'' in
  [{\em 2016 IEEE International Conference on Consumer Electronics-China
  (ICCE-China)}{\nolinebreak\hspace{0.1em}]},   1--4, IEEE (2016).

\bibitem{dyntex}
Ghanem, B. and Ahuja, N., ``Maximum margin distance learning for dynamic
  texture recognition,'' in [{\em European Conference on Computer
  Vision}{\nolinebreak\hspace{0.1em}]},   223--236, Springer (2010).

\bibitem{hamza2019parameter}
Hamza, A.~M., Abdelazim, A., and Ait-Boudaoud, D., ``Parameter optimization in
  {H. 265} rate-distortion by single frame semantic scene analysis,'' {\em
  Electronic Imaging}~{\bf 2019}(11),  262--1 (2019).

\bibitem{derf}
Xiph.org, ``{Xiph.org Video Test Media}.''
\newblock https://media.xiph.org/video/derf/.

\bibitem{zhang_bull}
{Zhang}, F. and {Bull}, D.~R., ``Rate-distortion optimization using adaptive
  lagrange multipliers,'' {\em IEEE Transactions on Circuits and Systems for
  Video Technology}~{\bf 29},  3121--3131 (Oct 2019).

\bibitem{Papadopoulos}
{Papadopoulos}, M.~A., {Zhang}, F., {Agrafiotis}, D., and {Bull}, D., ``An
  adaptive {QP} offset determination method for {HEVC},'' in [{\em 2016 IEEE
  International Conference on Image Processing
  (ICIP)}{\nolinebreak\hspace{0.1em}]},   4220--4224 (Sep. 2016).

\bibitem{yang2017perceptual}
Yang, A., Zeng, H., Chen, J., Zhu, J., and Cai, C., ``Perceptual feature guided
  rate distortion optimization for high efficiency video coding,'' {\em
  Multidimensional Systems and Signal Processing}~{\bf 28}(4),  1249--1266
  (2017).

\bibitem{LingICME}
{Ling}, S., {Baveye}, Y., {Callet}, P.~L., {Skinner}, J., and {Katsavounidis},
  I., ``Towards perceptually-optimized compression of user generated content
  (ugc): Prediction of ugc rate-distortion category,'' in [{\em 2020 IEEE
  International Conference on Multimedia and Expo
  (ICME)}{\nolinebreak\hspace{0.1em}]},   1--6 (2020).

\bibitem{EIRingis}
Ringis, D.~J., Pitie, F., and Kokaram, A., ``Per clip lagrangian multiplier
  optimisation for ({HEVC}),'' {\em Electronic Imaging}~{\bf 2020}(12) (2020).

\bibitem{bdrate}
Bjontegaard, G., ``Calculation of average {PSNR} differences between rd curves;
  {VCEG-M33},'' tech. rep. (2001).

\bibitem{numericalmethods}
Flannery, B.~P., Press, W.~H., Teukolsky, S.~A., and Vetterling, W.,
  ``Numerical recipes in {C},'' {\em Press Syndicate of the University of
  Cambridge, New York}~{\bf 24},  78 (1992).

\bibitem{netflixdb}
Netflix, ``{Netflix Open Content}.''
\newblock https://opencontent.netflix.com/.

\bibitem{MCL}
Lin, J.~Y., Jin, L., Hu, S., Katsavounidis, I., Li, Z., Aaron, A., and Kuo,
  C.-C.~J., ``Experimental design and analysis of jnd test on coded
  image/video,'' in [{\em Applications of digital image processing
  XXXVIII}{\nolinebreak\hspace{0.1em}]},   {\bf 9599},  95990Z, International
  Society for Optics and Photonics (2015).

\bibitem{videoLAN}
VideoLAN, ``{x265, the free {H.265} encoder}.''
\newblock https://www.videolan.org/developers/x265.html.

\bibitem{covell2016optimizing}
Covell, M., Arjovsky, M., Lin, Y.-C., and Kokaram, A., ``Optimizing transcoder
  quality targets using a neural network with an embedded bitrate model,'' {\em
  Electronic Imaging}~{\bf 2016}(2),  1--7 (2016).

\end{thebibliography}
\bibliographystyle{spiebib} % makes bibtex use spiebib.bst

\end{document}